\documentclass[floatfix,amsmath,amssymb,aps,prc,twocolumn,superscriptaddress,showpacs,preprintnumbers,nofootinbib]{revtex4-2}

\usepackage{graphicx,color}
\usepackage{multirow}
\usepackage{times}
\usepackage{bm}
\usepackage{epstopdf}
\usepackage{enumerate}
\usepackage{ulem}
\usepackage{hyperref}
\usepackage{cancel}

\newcommand{\MeV}{\textrm{MeV}}

\graphicspath{{./Figs/}}

\makeatletter
\def\@fnsymbol#1{\ensuremath{\ifcase#1\or \dagger\or \ddagger\or
   \mathsection\or \mathparagraph\or \|\or **\or \dagger\dagger
   \or \ddagger\ddagger \else\@ctrerr\fi}}
    \makeatother
\newcommand{\deceased}[1]{\altaffiliation{#1}}

\begin{document}

\title{
Repulsive $\Lambda$ potentials in dense neutron star matter and binding energy of $\Lambda$ in hypernuclei
}
\author{Asanosuke Jinno}
\affiliation{
Department of Physics, Faculty of Science,
Kyoto University, Kyoto 606-8502, Japan}

\author{Koichi Murase}
\affiliation{Yukawa Institute for Theoretical Physics,
Kyoto University, Kyoto 606-8502, Japan}

\author{Yasushi Nara}
\affiliation{
Akita International University, Yuwa, Akita-city 010-1292, Japan}

\author{Akira Ohnishi}\deceased{Deceased.}
\affiliation{Yukawa Institute for Theoretical Physics,
Kyoto University, Kyoto 606-8502, Japan}

\date{\today}

\preprint{YITP-23-78, KUNS-2968}
\begin{abstract}
The repulsive three-body force between the lambda ($\Lambda$) hyperon and medium nucleons
is a key element in solving the hyperon puzzle in neutron stars.
We investigate the binding energies of the $\Lambda$
hyperon in hypernuclei
to verify the repulsive $\Lambda$ potentials from
the chiral effective field theory ($\chi$EFT) employing
the Skyrme Hartree-Fock method.
We find that the $\chi$EFT $\Lambda$ potential with $\Lambda NN$ three-body forces reproduces
the existing hypernuclear binding energy data, whereas
the $\Lambda$ binding energies are overestimated without the $\Lambda NN$ three-body force.
Additionally,
we search for the parameter space of the $\Lambda$ potentials by varying the Taylor coefficients of the $\Lambda$ potential and the effective mass of $\Lambda$ at the saturation density.
Our analysis demonstrates that the parameter region consistent with the $\Lambda$ binding energy data spans a wide range of the parameter space, including even more repulsive potentials than the $\chi$EFT prediction.
We confirm that these strong repulsive $\Lambda$ potentials suppress the presence of $\Lambda$ in neutron star matter.
We found that the $\Lambda$ potentials repulsive at high densities are favored when the depth of the $\Lambda$ potential at the saturation density, $U_\Lambda(\rho_0)=J_\Lambda$, is $J_\Lambda\gtrsim-29~\MeV$, while attractive ones are favored when $J_\Lambda \lesssim -31~\MeV$.
This suggests that future high-resolution data of hypernuclei could rule out the scenario in which $\Lambda$s appear 
through the precise determination of
$J_\Lambda$ within the accuracy of $1~\text{MeV}$.
\end{abstract}

\maketitle

\section{Introduction}
Neutron stars are gravitationally bound objects made of cold, extremely dense, and strongly interacting matter, which has a rich phase structure of quantum chromodynamics (QCD)~\cite{Baym:2017whm}. They provide a unique cosmic laboratory for studying matter under extreme conditions.
The inner structure of neutron stars is one of the important subjects in astrophysics and nuclear physics.
Based on astrophysical observations and nuclear experiments, possibilities of various exotic states inside neutron stars have been theoretically discussed, including the admixture of hyperons~\cite{Glendenning:1991es, Schaffner:1995th, Balberg:1997yw, Baldo:1999rq, Vidana:2000ew, Fortin:2017cvt}
and the transition from hadron to quark matter~\cite{Ozel:2010bz,Weissenborn:2011qu,Bonanno:2011ch,Klahn:2013kga,Baym:2017whm, Annala:2019puf, Kojo:2021wax}.

In the 20th century, hyperons were predicted to admix in the neutron star matter at a density of
$2$--$4\,\rho_0$, with the saturation density being $\rho_0\simeq 0.16~\text{fm}^{-3}$,
from phenomenological models
~\cite{Glendenning:1991es, Schaffner:1995th,Balberg:1997yw,Baldo:1999rq, Vidana:2000ew}
based on experimental data such as hypernuclear spectroscopy.
The admixture of the hyperons
softens the equation of state (EOS) of neutron star matter
and reduces the maximum allowed mass of neutron stars significantly.
For this reason, hyperonic matter EOSs 
constrained by hypernuclear data~\cite{Glendenning:1991es, Balberg:1997yw, Dapo:2008au}
or $G$-matrix calculations using two-body hyperon-nucleon ($YN$) interactions~\cite{Baldo:1999rq, Vidana:2000ew, Nishizaki:2002ih}
could not sustain the observed massive neutron stars
with $M \gtrsim 2 M_\odot$~\cite{Demorest:2010bx, Antoniadis:2013pzd, Fonseca:2016tux, NANOGrav:2019jur, Miller:2021qha}.
This problem is known as the hyperon puzzle in neutron stars
and has been attracting the attention of nuclear physics and astrophysics researchers.
A number of possible scenarios to solve the hyperon puzzle have been proposed, such as
repulsive hyperon potentials at high densities
caused by many-body baryon interactions~\cite{Nishizaki:2002ih, Lonardoni:2014bwa, Nagels:2015lfa, Gerstung:2020ktv, Yamamoto:2013ada, Yamamoto:2014jga, Nagels:2015lfa, Yamamoto:2015lwa, Togashi:2016fky, Haidenbauer:2016vfq, Logoteta:2019utx, Choi:2021xdm},
hyperon-hyperon ($YY$) repulsion
~\cite{Weissenborn:2011ut, Chatterjee:2015pua, Fortin:2017cvt},
and a continuous transition to quark matter before the hyperon admixture~\cite{Baym:2017whm, Kojo:2021wax},
yet the definitive answer to the puzzle has not been found so far.

In the following, we shall explore the
first scenario: the repulsive hyperon potential
at high densities caused by many-baryon interactions.
The three-nucleon repulsion is known to be necessary
to explain the nuclear matter saturation point.
Therefore, it is worthwhile to examine the impact of the $YNN$ ($YYN$, $YYY$) three-baryon repulsion on the neutron star matter EOS\@.
Several models including many-body interactions have been investigated by
a multi-Pomeron exchange potential (MPP) 
based on the extended soft core (ESC) model~\cite{Yamamoto:2013ada, Yamamoto:2014jga, Nagels:2015lfa, Yamamoto:2015lwa}
and the KIDS (Korea-IBS-Daegu-SKKU) density functional formalism~\cite{Choi:2021xdm}.
These results show that
repulsive three-baryon forces, which reproduce
the binding energies of $\Lambda$ hypernuclei, may solve the hyperon puzzle.

The chiral effective field theory ($\chi$EFT) provides a systematic and model-independent approximation of QCD at low densities~\cite{Epelbaum:2008ga}.
The three-body forces were found to cause repulsive $\Lambda$ potential
at high densities in the $\chi$EFT~\cite{Gerstung:2020ktv, Kohno:2018gby, Logoteta:2019utx, Haidenbauer:2016vfq}
with the decuplet saturation model for the three-baryon interactions~\cite{Petschauer:2016pbn}.
This $\Lambda$ potential, $U^\mathrm{chi3}_\Lambda$,
can prevent $\Lambda$ from appearing in neutron stars~\cite{Gerstung:2020ktv}
so that we can avoid the hyperon puzzle.
The proposed $\Lambda$ potential $U^\mathrm{chi3}_\Lambda$ should be validated by experiments and observations.
In particular, the density dependence~\cite{Gerstung:2020ktv, Kohno:2018gby} and the momentum dependence~\cite{Kohno:2018gby}
of the $\Lambda$ potential $U^\mathrm{chi3}_\Lambda$
can be tested
with different types of experimental data,
such as the observables from heavy-ion collisions and the $\Lambda$ binding energies of hypernuclei.

Heavy-ion collision experiments provide a unique
opportunity to study the properties of QCD matter under various densities and temperatures.
In the past two decades,
a quark-gluon plasma of high temperature and low baryon density
has been created and actively studied~\cite{Akiba:2015jwa,  Arslandok:2023utm}
using heavy-ion collisions at the top energies
of the Relativistic Heavy Ion Collider (RHIC) and the Large Hadron Collidier (LHC).
In recent heavy-ion experiments at lower energies, such as RHIC Beam Energy Scan~\cite{Luo:2015doi} and NA61/SHINE programs~\cite{NA61:2014lfx}, 
two colliding nuclei are compressed to form high baryon-density matter.
In this regime, the anisotropic flows are known to be sensitive to the hadronic potentials at high densities~\cite{Danielewicz:2002pu}.
The transport model calculation~\cite{Nara:2021fuu,Nara:2022kbb} using the event generator JAM2~\cite{JAM2}
with the $\Lambda$ potential
$U^\mathrm{chi3}_\Lambda$ 
explain the collision energy dependence of the proton and $\Lambda$ directed flow slopes ($dv_1/dy$) in Au + Au collisions
within a relative precision of around 20\%
in the collision energy range of
$3 \leq \sqrt{s_{NN}} \leq 19.6~\text{GeV}$%
~\cite{STAR:2017okv,STAR:2020dav,STAR:2021yiu}.
However, it is found in Ref.~\cite{Nara:2022kbb} that
the directed flow of $\Lambda$ is not very sensitive
to the density dependence of the $\Lambda$ potential
while it is sensitive to the momentum dependence.
We should investigate further observables sensitive to the density dependence.

As another experimental observable to test  $U^\mathrm{chi3}_\Lambda$,
the up-to-date data of the $\Lambda$ binding energies in hypernuclei provide useful constraints.
The density dependence of the $\Lambda$ potential at lower densities $\rho<\rho_0$ can be constrained
by the $\Lambda$ binding energies~\cite{Millener:1988hp}
while the momentum dependence is sensitive to the energy difference between orbitals of the $\Lambda$ binding energy through the effective mass~\cite{Yamamoto:1988qz}.
The $\Lambda$ potentials LY-IV~\cite{Lanskoy:1997xq} and HP$\Lambda$2~\cite{Guleria:2011kk} reproduce the $\Lambda$ binding energy for hypernuclei in a wide range of mass number.
Millner \textit{et al.}~\cite{Millener:1988hp} pointed out that
the density dependence at higher densities determined by the best fitting to the $\Lambda$ binding energy largely depends on the fitting form
while the density dependence at lower densities is well constrained.
However, the $\Lambda$ potential $U^\mathrm{chi3}_\Lambda$ has a distinct shape from LY-IV and HP$\Lambda$2:
it is more attractive at $\rho<\rho_0$ and has a larger value of the effective mass than LY-IV and HP$\Lambda$2.
Thus, it should be tested whether $U^\mathrm{chi3}_\Lambda$ can reproduce the $\Lambda$ binding energy data.
In Ref.~\cite{Haidenbauer:2019thx},
the $\Lambda$ potential based on
the next-to-leading order (NLO) $\chi$EFT~\cite{Haidenbauer:2013oca,Haidenbauer:2019boi}
that includes
only two-body interaction,
has been tested by the $G$-matrix calculation.
This two-body interaction only partially reproduces the observed binding energies for hypernuclei with mass number $A > 12$.

In this paper,
we show that the $\Lambda$ potential $U^\mathrm{chi3}_\Lambda$ from the $\chi$EFT~\cite{Gerstung:2020ktv, Kohno:2018gby}, which is repulsive at high densities, can reproduce the experimental data of the $\Lambda$ binding energies with
the Skyrme-Hartree-Fock method~\cite{Rayet:1976fs, Rayet:1981uj, Fernandez:1989zza, Lanskoy:1997xq, Guleria:2011kk}.
These results mean that the two distinct $\Lambda$ potentials---repulsive and attractive $\Lambda$ potentials at high densities---can reproduce the same experimental $\Lambda$ binding energy.
Here, we carry out the global parameter search for the $\Lambda$ potentials to scan a wider range of the $\Lambda$ potentials and evaluate the uncertainty range.
We parametrize the $\Lambda$ potential by the effective mass and the Taylor coefficients at $\rho_0$.
We examine different $\Lambda$ potentials by varying the parameters
and identify the parameter space consistent with the experimental data.
Finally,
we discuss the parameter region of the $\Lambda$ potentials which suppresses $\Lambda$'s in neutron star matter.

This paper is organized as follows.
In Sec.~\ref{sec:SkyrmeLamPot}, we introduce the Skyrme-Hartree-Fock method for the $\Lambda$ hypernuclei with the $\Lambda$ potential from the $\chi$EFT and compare the $\Lambda$ binding energies with existing models and the experimental data.
In Sec.~\ref{sec:ModelIndep}, we search for the favored region of the effective mass and the Taylor coefficients of the $\Lambda$ potential.
In Sec.~\ref{sec:LamAdmix},
we discuss the admixture of $\Lambda$ in neutron star matter for various $\Lambda$ potentials which reproduce the binding energy of $\Lambda$ hypernuclei.
The conclusion and outlook are given in Sec.~\ref{sec:summary}.

\section{$\Lambda$ binding energy from the $\chi$EFT}
\label{sec:SkyrmeLamPot}

To explain the $\Lambda$ binding energies of $\Lambda$ hypernuclei from middle to large mass numbers,
mean-field calculations of self-consistent calculations
have been successfully employed,
including Skyrme-Hartree-Fock methods~\cite{Rayet:1976fs,Rayet:1981uj,Fernandez:1989zza,Lanskoy:1997xq,Win:2010tq,Guleria:2011kk,Choi:2021xdm},
relativistic mean-field models~\cite{Win:2008vw,Tsubakihara:2009zb,Tanimura:2011ji,Fortin:2017cvt}, and $G$-matrix calculations~\cite{Nagels:2015lfa,Logoteta:2019utx}.

We employ the Skyrme-Hartree-Fock method to compute
the binding energy of $\Lambda$ hypernuclei
using the Skyrme-type $\Lambda$ potential parametrizing the results of the $\chi$EFT\@.

\subsection{Skyrme-Hartree-Fock method for $\Lambda$ hypernuclei}
\label{sec:SHF-method}

In this study, the wave function of the $\Lambda$ hypernuclei ${}^{A}_\Lambda Z$ with mass number $A$ and proton number $Z$ is taken as~\cite{Rayet:1976fs}
\begin{equation}
    \label{eq:wavefunc}
     \Phi_\mathrm{hyp} = \phi_\Lambda \Phi_\mathrm{core},
\end{equation}
where $\phi_\Lambda(\bm{r}, \sigma)$ is the single-particle wave function of $\Lambda$, with $\bm{r}$ and $\sigma=\pm1/2$ being the spatial and spin coordinates, respectively.
The Slater determinant $\Phi_\text{core}$ is constructed from $A-1$ nucleon single-particle wave functions $\phi_i(\bm{r}, q, \sigma)$ with $q=\pm1/2$ being the isospin.
The Skyrme-type baryon-baryon interaction with one $\Lambda$ hyperon is expressed as
\begin{align}
    \label{eq:int}
    V =& V^{NN} + V^{\Lambda N} \nonumber \\
    =& \sum_{i<j}^{A-1} v^{NN}(\bm{r}_i,\bm{r}_j)
    + \sum_{i=1}^{A-1} v^{\Lambda N}(\bm{r}_\Lambda,\bm{r}_i),
\end{align}
where $\bm{r}_i$ and $\bm{r}_\Lambda$ are the spatial coordinates of the $i$th nucleon and $\Lambda$, respectively.
We use the SLy4 parametrization~\cite{Chabanat:1997un} for the nucleon-nucleon interaction $v^{NN}$.
The following $\Lambda$--nucleon interaction $v^{\Lambda N}$ is assumed as in Refs.~\cite{Rayet:1981uj, Lanskoy:1997xq, Choi:2021xdm}:
\begin{align}
    &v^{\Lambda N}(\bm{r}_\Lambda , \bm{r}_N) \nonumber \\
    &\quad =
    t^\Lambda_0 (1 + x^\Lambda_0 P_\sigma) \delta(\bm{r}_\Lambda - \bm{r}_N) \nonumber \\
    &\qquad +\dfrac{1}{2} t^\Lambda_1 \bigl[\overleftarrow{\bm{k}}^2 \delta(\bm{r}_\Lambda - \bm{r}_N) + \delta(\bm{r}_\Lambda - \bm{r}_N) \overrightarrow{\bm{k}}^2\bigr] \nonumber \\
    &\qquad + t^\Lambda_2 \overleftarrow{\bm{k}} \delta(\bm{r}_\Lambda - \bm{r}_N) \cdot \overrightarrow{\bm{k}} \nonumber \\
    &\qquad +\dfrac{3}{8} t^\Lambda_{3,1} (1+x^\Lambda_{3,1} P_\sigma) \delta(\bm{r}_\Lambda - \bm{r}_N) \rho_N^{\gamma_1} \left(\dfrac{\bm{r}_N+\bm{r}_\Lambda}{2}\right) \nonumber \\
    &\qquad + \dfrac{3}{8} t^\Lambda_{3,2} (1+x^\Lambda_{3,2} P_\sigma) \delta(\bm{r}_\Lambda - \bm{r}_N) \rho_N^{\gamma_2} \left(\dfrac{\bm{r}_N+\bm{r}_\Lambda}{2}\right),
    \label{eq:LN}
\end{align}
where $t^\Lambda_k$, $t^\Lambda_{k,l}$, $x^\Lambda_{k}$, and $x^\Lambda_{k,l}$ are the Skyrme potetial parameters. The spin-exchange operator is given by $P_\sigma=(1+\bm{\sigma}_\Lambda \cdot \bm{\sigma}_N)/2$ with $\bm{\sigma}_\Lambda$ and $\bm{\sigma}_N$ being the Pauli matrices acting on the spin wave functions of $\Lambda$ and the nucleon, respectively. The derivatives, $\overleftarrow{\bm{k}}=-(\overleftarrow{\nabla}_\Lambda-\overleftarrow{\nabla}_N)/2i$ and $\overrightarrow{\bm{k}}=(\overrightarrow{\nabla}_\Lambda-\overrightarrow{\nabla}_N)/2i$, operate on the left- and right-hand sides, respectively, where $\nabla_\Lambda = \partial/\partial \bm{r}_\Lambda$, and $\nabla_N = \partial/\partial \bm{r}_N$. The symbol $\rho_N(\bm{r}_N)$ denotes the nucleon density specified later.
The $\rho_N^{\gamma_1}$ term with $\gamma_1=1/3$, which is motivated by the expansion in the Fermi momentum ($\propto \rho_N^{1/3}$), is needed for the Skyrme-type $\Lambda$ potential to reproduce the $G$-matrix result of the $\Lambda$ potential~\cite{Lanskoy:1997xq}.
The $\rho_N^{\gamma_2}$ term with $\gamma_2=2/3$ is added to reproduce the results from the $\chi$EFT~\cite{Gerstung:2020ktv}.
The spin-orbit force is neglected in this work because it is expected to be small from the experimental data~\cite{Ajimura:2001na,Akikawa:2002tm}.
The pairing correlation is not considered either.

The expectation value of the total energy for the $\Lambda$ hypernuclei $\mathcal{E}_\mathrm{hyp}$ is obtained as
\begin{align}
    \mathcal{E}_\mathrm{hyp} &=
    \left\langle \Phi_\mathrm{hyp} \right\vert T + V \left\vert \Phi_\mathrm{hyp} \right\rangle
    - \mathcal{E}_\mathrm{c.m.} \nonumber \\
    &= \mathcal{E}_N + \mathcal{E}_\Lambda - \mathcal{E}_\mathrm{c.m.},
    \label{eq:shf-hypernucleus-energy}
\end{align}
where
\begin{equation}
    T= T_N + T_\Lambda = -\sum_{i=1}^{A-1} \dfrac{\hbar^2\nabla_i^2}{2m_i} - \dfrac{\hbar^2\nabla_\Lambda^2}{2 m_\Lambda}
\end{equation}
is the kinetic energy operator,
and $m_i$ and $m_\Lambda$ are the masses of the nucleon and $\Lambda$, respectively.
The energies $\mathcal{E}_N$ and $\mathcal{E}_\Lambda$ are
contributions from $T_N + V^{NN}$ and $T_\Lambda + V^{\Lambda N}$, respectively.
The total energy of $\Lambda$
takes the following form
\begin{align}
\mathcal{E}_\Lambda
=&\int \mathrm{d}^3 r H_\Lambda, \\
H_\Lambda=& \dfrac{\hbar^2}{2 m_\Lambda} \tau_\Lambda + a^\Lambda_1 \rho_N \rho_\Lambda \nonumber \\
    &+ a^\Lambda_2 (\tau_\Lambda \rho_N + \tau_N \rho_\Lambda) - a^\Lambda_3 (\rho_\Lambda  \cdot \Delta \rho_N) \nonumber \\
    &+ a_4^\Lambda \rho_N^{1+\gamma_1} \rho_\Lambda
    + a_5^\Lambda \rho_N^{1+\gamma_2} \rho_\Lambda,
    \label{eq:HLambda}
\end{align}
where $H_\Lambda$ is the energy density of $\Lambda$, and
the coefficients $a^\Lambda_i$ are related to the parameters in the $\Lambda$-nucleon interaction $v^{\Lambda N}$ as
\begin{align}
    a_1^\Lambda &= t^\Lambda_0 \left(1+\dfrac{1}{2}x^\Lambda_0\right), &
    a_2^\Lambda &= \dfrac{1}{4}(t^\Lambda_1+t^\Lambda_2), \nonumber \\
    a_3^\Lambda &= \dfrac{1}{8}(3t^\Lambda_1-t^\Lambda_2), &
    a_4^\Lambda &= \dfrac{3}{8}t^\Lambda_{3,1}\left(1+\dfrac{1}{2}x^\Lambda_{3,1}\right), \nonumber \\
    a_5^\Lambda &= \dfrac{3}{8}t^\Lambda_{3,2}\left(1+\dfrac{1}{2}x^\Lambda_{3,2}\right).\nonumber
\end{align}
The densities $\rho_N$, $\tau_N$, $\rho_\Lambda$, and $\tau_\Lambda$ are defined as
\begin{align}
    \label{eq:Ndensity}
    \rho_N &= \sum_{i=1}^{A-1} |\phi_i|^2, &
    \tau_N &= \sum_{i=1}^{A-1} |\nabla \phi_i|^2, \\
    \label{eq:Ldensity}
    \rho_\Lambda &= |\phi_\Lambda|^2, &
    \tau_\Lambda &= |\nabla \phi_\Lambda|^2.
\end{align}
The center-of-mass energy $\mathcal{E}_\mathrm{c.m.}$ in Eq.~\eqref{eq:shf-hypernucleus-energy} is approximated by the average of the center-of-mass kinetic operator neglecting the cross terms~\cite{Rayet:1976fs},
\begin{equation}
    \mathcal{E}_\mathrm{c.m.} \simeq
    \int \mathrm{d}^3 r
    \dfrac{\hbar^2 \left(\tau_N +  \tau_\Lambda\right)}{2\left[Z m_p + (A-Z-1) m_n + m_\Lambda\right]}.
\end{equation}

The Hartree-Fock equations for single-particle wave functions $\phi_i$ ($i=1,2,\cdots,A-1,\Lambda$) are derived from the variational equation:
\begin{equation}
    \label{eq:HFeq}
    \dfrac{\delta}{\delta {\phi_i}} \left(\mathcal{E}_\mathrm{hyp} - \sum_{i=1}^{A-1} \epsilon_i \int \mathrm{d}^3{r} \left| \phi_i \right|^2 - \epsilon_\Lambda \int \mathrm{d}^3 r \left| \phi_\Lambda \right|^2 \right) = 0,
\end{equation}
where $\epsilon_i$ is the single-particle energy. 
Equation~\eqref{eq:HFeq} combined with Eq.~\eqref{eq:shf-hypernucleus-energy} yields the equation for the single-particle wave function of the $i$th particle with the baryon type $B_i=p,n,\Lambda$,
\begin{multline}
\label{eq:SHFeq}
    \Bigg[ - \nabla \cdot \left(\dfrac{\hbar^2}{2 m_{B_i}^*(\bm{r})} \nabla\right) + V_{B_i}(\bm{r}) \\
    - i \bm{W}_{B_i}(\bm{r}) \cdot (\nabla \times {\bm{\sigma}}) \Bigg] \phi_i = \epsilon_i \phi_i.
\end{multline}
The first term is the kinetic energy including the effective mass, the second is the single-particle potential, and the third is the spin-orbit potential.
The expressions for 
those terms
are specified below.

In this work, we assume spherical symmetry of the hypernuclei as in Refs.~\cite{Rayet:1976fs, Rayet:1981uj, Fernandez:1989zza, Lanskoy:1997xq, Guleria:2011kk, Choi:2021xdm}.
We assign the principle quantum number $n$, the orbital angular momentum $\ell$, the total angular momentum $j$, and the magnetic quantum number $m_j$ to each $i$.
The single-particle wave function for the $i$th nucleon with the isospin $q$ is expressed as 
\begin{align}
    \phi_i({\bm r},q)&=\dfrac{R_\alpha(r)}{r} \mathcal{Y}_{ljm_j}(\hat{\bm{r}}) \chi_q, &
    \alpha&=\{nljq\},
\end{align}
where $r=|\bm{r}|$, $\hat{\bm{r}} = \bm{r}/|\bm{r}|$, and
\begin{align}
    \mathcal{Y}_{ljm_j}(\hat{\bm{r}})=\sum_{m_l m_s} \langle l m_l~1/2~m_s|j m_j \rangle Y_{l m_l}(\hat{\bm r}) \chi_{m_s}.
\end{align}
The symbols $\chi_q$ and $\chi_{m_s}$ denote
the isospin and spin wave functions, respectively,
and $Y_{l m_l}$ is the spherical harmonics.
The $\Lambda$ single-particle wave function is similarly written as
\begin{align}
    \phi_\Lambda({\bm r})&=\dfrac{R_\alpha(r)}{r} \mathcal{Y}_{ljm_j}(\hat{\bm{r}}).
\end{align}
The Skyrme-Hartree-Fock equation~\eqref{eq:SHFeq} is reduced to the equation for the radial wave function $R_\alpha$:
\begin{multline}
-\dfrac{\hbar^2}{2 m_{B_\alpha}^*}R''_\alpha(r) -\dfrac{\mathrm{d}}{\mathrm{d}r}\left(\dfrac{\hbar^2}{2 m_{B_\alpha}^*}\right) R'_\alpha (r) \\
+\Bigg[ \dfrac{\hbar^2}{2 m_{B_\alpha}^*} \dfrac{l_\alpha (l_\alpha+1)}{r^2} +V_{B_\alpha}(r) + \dfrac{1}{r} \dfrac{\mathrm{d}}{\mathrm{d}r}
\left(\dfrac{\hbar^2}{2m_{B_\alpha}^*}\right) \\
+\dfrac{W_{B_\alpha}}{r}\left(j_\alpha(j_\alpha+1)-l_\alpha(l_\alpha+1) - \dfrac{3}{4}\right) \Bigg]R_\alpha(r) \\
= \epsilon_\alpha R_\alpha(r).
\end{multline}
For nucleons, the explicit forms of $m^*_{B_\alpha}$, $V_{B_\alpha}$, and $W_{B_\alpha}$ are found in Appendix~\ref{app:expressions}\@. For $\Lambda$, they are expressed as
\begin{align}
    \dfrac{\hbar^2}{2m_\Lambda^*} =& \dfrac{\hbar^2}{2 m_\Lambda} + a^\Lambda_2 \rho_N, \\
    V_\Lambda =&  a^\Lambda_1\rho_N + a^\Lambda_2 \tau_N - a^\Lambda_3 \Delta{\rho_N} \nonumber \\
    &+ a^\Lambda_4 {\rho_N}^{1+\gamma_1}
    + a^\Lambda_5 {\rho_N}^{1+\gamma_2}. \label{eq:VLambda}
\end{align}
In the case of the open-shell nuclei,
we employ the filling approximation~\cite{Perez-Martin:2008dlm}:
when there are $m$ nucleons in the open shell, they are filled in the highest $j$ states at the same occupation probability $m/(2j+1)$.
Similarly, $\Lambda$ occupies each state of $j$ with an occupation probability $1/(2j+1)$.
Then, the $\Lambda$ density is calculated as
\begin{equation}
    \rho_\Lambda(r)=\dfrac{1}{2j+1}\sum_{m_j}\left|\dfrac{R_\alpha(r)}{r} \mathcal{Y}_{ljm_j} \right|^2 = \dfrac{R^2_\alpha(r)}{4\pi r^2}.
\end{equation}

After solving Eq.~\eqref{eq:SHFeq} self-consistently,
the $\Lambda$ binding energy is obtained as
\begin{equation}
    B_\Lambda=-(\mathcal{E}_\mathrm{hyp} - \mathcal{E}_\mathrm{core}),
\end{equation}
where the total energy of the core nucleus $\mathcal{E}_\mathrm{core}$ is independently calculated by solving the self-consistent equation for the Slater determinant of the core nucleus.

\subsection{Skyrme-type $\Lambda$ potentials from the $\chi$EFT}

In this subsection, we parametrize the $\Lambda$ potentials obtained from the $\chi$EFT assuming the form of the density and momentum dependence
in Eq.~\eqref{eq:HLambda}.
In uniform matter, the kinetic density $\tau_\Lambda$ becomes
\begin{equation}
    \tau_\Lambda = |\nabla \phi_\Lambda|^2 = k_\Lambda^2 |\phi_\Lambda|^2 = k_\Lambda^2\rho_\Lambda,
\end{equation}
with $k_\Lambda$ being the momentum of $\Lambda$. Then, the $\Lambda$ single-particle potential in uniform nuclear matter at zero temperature is obtained as
\begin{multline}
U_{\Lambda}(\rho_N, k_\Lambda)
    = \dfrac{\delta}{\delta \rho_\Lambda} \left[H_\Lambda - \dfrac{\hbar^2}{2 m_\Lambda} \tau_\Lambda\right] \\
    = a^\Lambda_1 \rho_N
    + a^\Lambda_2(k_\Lambda^2 \rho_N+ \tau_N)
    + a^\Lambda_4 \rho^{1+\gamma_1}_N
    + a^\Lambda_5 \rho^{1+\gamma_2}_N,
    \label{eq:PotLamSkyrmeUnif}
\end{multline}
with
\begin{align}
    \tau_{N} &= \tau_{p} + \tau_{n}, \\
    \tau_{q} &= \dfrac{3}{5} \left({3 \pi^2}\right)^{2/3} \rho_q^{5/3}, \quad q=p,n, \\
    \gamma_1 & = 1/3, \quad\gamma_2=2/3.
\end{align}
The Skyrme potential parameters of $\Lambda$, $a^\Lambda_i$, are determined by the fitting to the potentials from the $\chi$EFT of Ref.~\cite{Gerstung:2020ktv} (GKW)
and Ref.~\cite{Kohno:2018gby} (Kohno).

\begin{figure}[tbhp]
\includegraphics[width=8cm]{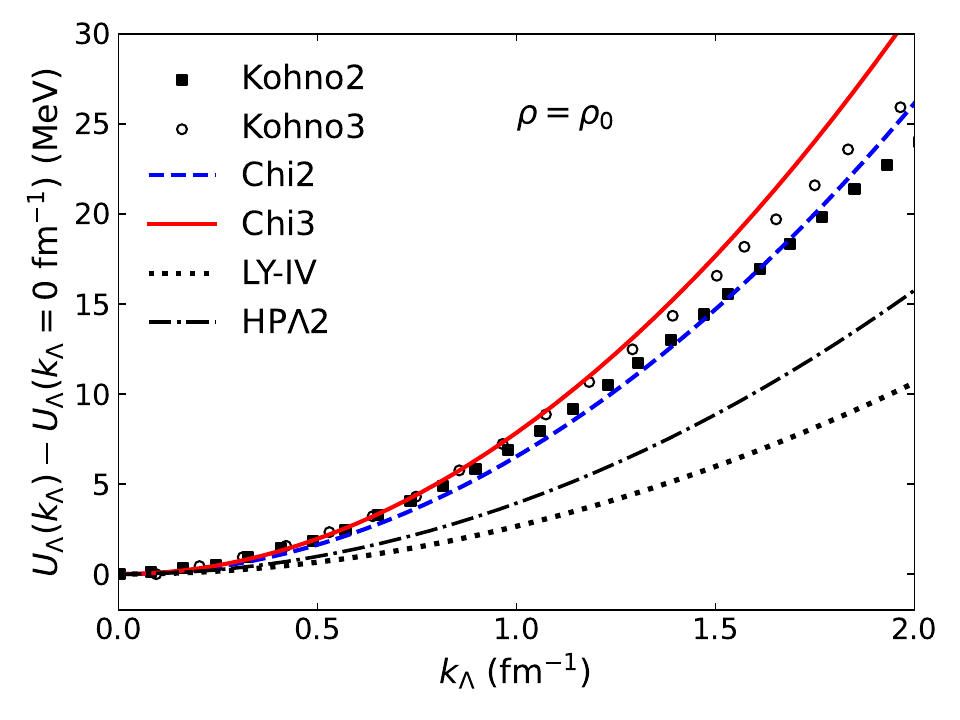}
\caption{Momentum dependence of the $\Lambda$ potentials in symmetric nuclear matter at the saturation density with its value at zero momentum subtracted. 
Kohno2 and Kohno3 represent the results of the $\Lambda$ single-particle potential
with only two-body interactions
and two- and three-body interactions~\cite{Kohno:2018gby} from the $\chi$EFT~\cite{Kohno:2018gby}, respectively.
Solid and dashed lines represent the fitting results to Kohno2 and Kohno3, respectively.
The dotted and dash-dotted lines correspond to the $\Lambda$ potentials 
LY-IV~\cite{Lanskoy:1997xq}
and HP$\Lambda$2~\cite{Guleria:2011kk}, respectively.
}
\label{fig:MomDep}
\end{figure}

We first fix the momentum-dependent part of
the $\Lambda$ potential $a_2^\Lambda$ at the saturation density
$\rho_0=0.16\ \text{fm}^{-3}$ by using the results of
Ref.~\cite{Kohno:2018gby}.
Figure~\ref{fig:MomDep} shows the momentum dependence of the $\Lambda$ single-particle potential.
Kohno3 in Fig.~\ref{fig:MomDep}
is obtained by including
the $\Lambda NN$ three-body interaction using the decuplet saturation model~\cite{Petschauer:2016pbn}.
Kohno2 in Fig.~\ref{fig:MomDep} is obtained
using only the two-body force.
Because the depth of the $\Lambda$ potential at the saturation density differs between GKW~\cite{Gerstung:2020ktv} and Kohno~\cite{Kohno:2018gby}, 
we subtract the value at $k_\Lambda=0~\text{fm}^{-1}$ from the $\Lambda$ potential as $U_\Lambda(k_\Lambda)-U_\Lambda(k_\Lambda=0\ \text{fm}^{-1})$
and use it in fitting the momentum dependence of the $\Lambda$ potential.
Chi2 and Chi3 in Fig.~\ref{fig:MomDep} are the fitting results of Kohno2 and Kohno3 at $k_\Lambda<1.5~\text{fm}$, respectively. 

\begin{figure}[tbhp]
\includegraphics[width=8cm]{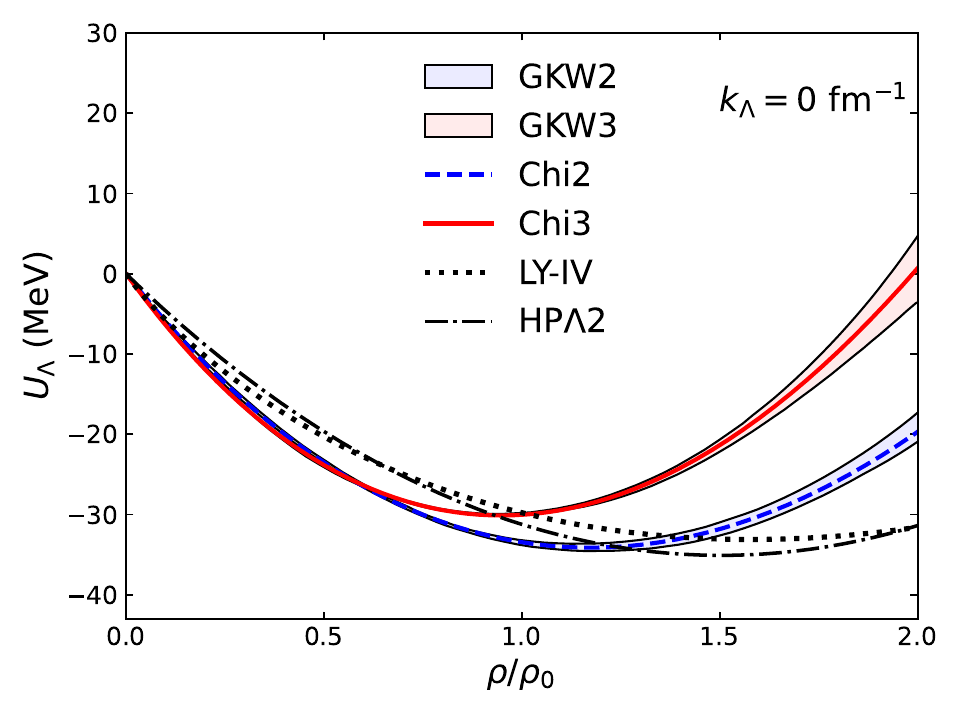}
\caption{Normalized baryon density dependence of the single-particle potentials for $\Lambda$ in symmetric nuclear matter.
GKW2 and GKW3 represent the results of the $\Lambda$ single-particle potential with only two-body interactions and with two- and three-body interactions obtained from the $\chi$EFT~\cite{Gerstung:2020ktv}, respectively.
The solid and dashed lines represent the fitting results to GKW2 and GKW3, respectively.
The dotted and dash-dotted lines correspond to the
$\Lambda$ potentials LY-IV~\cite{Lanskoy:1997xq}
and HP$\Lambda$2~\cite{Guleria:2011kk}, respectively.
}
\label{fig:DenseDep}
\end{figure}

For the remaining Skyrme potential parameters,
we consider the density dependence of the $\Lambda$ potential obtained from the $\chi$EFT~\cite{Gerstung:2020ktv}.
In Fig.~\ref{fig:DenseDep},
GKW2 (GKW3) is the result from the $\chi$EFT with $\Lambda N$ ($\Lambda N +\Lambda NN$) interaction. The $\Lambda NN$ three-body interaction in GKW3 is calculated by the decuplet saturation model~\cite{Petschauer:2016pbn}.
The parameters $(a^\Lambda_1, a^\Lambda_4, a^\Lambda_5)$ of Chi2 (Chi3) are obtained by fitting Eq.~\eqref{eq:PotLamSkyrmeUnif} to the upper and lower lines of GKW2 (GKW3) and taking the average.

Unlike the previous work~\cite{Nara:2022kbb}, where we fitted the data in the region $\rho/\rho_0<3.5$, the fitting region in the present study is limited to $\rho/\rho_0<1.5$ in order to reproduce the GKW results in the low-density region more accurately.
The resultant Skyrme parameters $a^\Lambda_1$, $a^\Lambda_2$,$a^\Lambda_4$, and $a^\Lambda_5$ from the fitting are listed in Table~\ref{tab:SkyrmeParams}.
The fitting results accurately reproduce the $\chi$EFT results at $\rho/\rho_0\lesssim1.0$, which is relevant for calculating the $\Lambda$ hypernuclei.

For comparison, the $\Lambda$ potentials that are used to explain the $\Lambda$ binding energy data (LY-IV~\cite{Lanskoy:1997xq} and HP$\Lambda$2~\cite{Guleria:2011kk}) are shown in Figs.~\ref{fig:MomDep} and~\ref{fig:DenseDep}. 
The LY-IV and HP$\Lambda$2 $\Lambda$ potentials exhibit different characteristics compared to the potentials obtained from the $\chi$EFT\@.
Specifically, they display enhanced repulsion in low-density regions, increased attraction in high-density regions, 
and a weaker momentum dependence.

Because the Skyrme parameter $a^\Lambda_3$ cannot be determined by fitting the results in the uniform matter, we determine $a^\Lambda_3$ to reproduce the experimental value of the $\Lambda$ binding energy of $^{13}_\Lambda {\rm C}$, $11.88~\text{MeV}$\@. The experimental value is
 taken from Ref.~\cite{Hashimoto:2006aw} with a correction of $0.5~{\text{MeV}}$, which is pointed out in Ref.~\cite{Gogami:2015tvu}. There are two reasons for choosing $^{13}_\Lambda {\rm C}$: First, it has a larger surface-energy effect compared with a heavier nucleus. Second, 
the spherical Skyrme-Hartree-Fock method is expected to provide a relatively good description of $^{13}_\Lambda {\rm C}$
because it has even numbers of protons and neutrons.

\begin{table}[tbhp]
\caption{The sets of Skyrme potential parameters
are listed above the gap between $\gamma_2$ and $J_\Lambda$.
Chi2 and Chi3 are the fitting results to the $\chi$EFT calculations~\cite{Gerstung:2020ktv,Kohno:2018gby}. 
LY-IV~\cite{Lanskoy:1997xq} and HP$\Lambda$2~\cite{Guleria:2011kk} are the $\Lambda$ potentials, which can explain the $\Lambda$ binding energy data.
We also list the values that characterize potentials:
the Taylor coefficients ($J_\Lambda$, $L_\Lambda$, $K_\Lambda$) and the normalized effective mass $m^*_\Lambda/m_\Lambda$ at $\rho_0$ defined by Eqs.~\eqref{eq:taylor1}--\eqref{eq:meff}.
The mean squared deviation of the calculated $\Lambda$ binding energy from the experimental data $\Delta B_\Lambda$ is defined by Eq.~\eqref{eq:RMSD}.
}
\centering
\begin{tabular}{l@{\qquad}rrrrrrr}
\hline
\hline
& Chi2 & Chi3 & LY-IV & HP$\Lambda$2 \\
\hline
$t_0^\Lambda~({\text{MeV}}~{\text{fm}^3})$ & $-352.2$ & $-388.3$ & $-542.5$ & $-399.9$\\
$t_1^\Lambda~({{\text{MeV}}~\text{fm}^5})$ & $143.7$ & $120.4$ & $56.0$ & $83.4$\\
$t_2^\Lambda~({{\text{MeV}}~\text{fm}^5})$ & $13.7$ & $68.7$ & $8.0$ & $11.5$\\
$t_{3,1}^\Lambda~({{\text{MeV}}~\text{fm}}^4)$ & $-951.9$ & $-1081.8$ & $1387.9$ & $2046.8$\\
$t_{3,2}^\Lambda~({{\text{MeV}}~\text{fm}}^5)$ & $2669$ & $3351$ & $0$ & $0$\\
$x_0^\Lambda$ & $0$ & $0$ & $-0.153$ & $-0.486$\\
$x_{3,1}^\Lambda$ & $0$ & $0$ & $0.107$ & $-0.660$\\
$x_{3,2}^\Lambda$ & $0$ & $0$ & $0$ & $0$\\
$\gamma_1$ & $1/3$ & $1/3$ & $1/3$ & $1$\\
$\gamma_2$ & $2/3$ & $2/3$ & $0$ & $0$\\
\hline
$J_\Lambda~\text{(MeV)}$ & $-33.5$ & $-30.0$ & $-29.8$ & $-31.2$\\
$L_\Lambda~\text{(MeV)}$ & $-23.5$ & $9.3$ & $-36.2$ & $-46.1$\\
$K_\Lambda~\text{(MeV)}$ & $415$ & $532$ & $218$ & $277$\\
$m^*_\Lambda/m_\Lambda$ & $0.73$ & $0.70$ & $0.87$ & $0.82$\\
$\Delta B_\Lambda~\text{(MeV)}$ & $1.55$ & $0.72$ & $0.71$ & $0.78$\\
\hline
\hline
\end{tabular}
\label{tab:SkyrmeParams}
\end{table}

We show in Table~\ref{tab:SkyrmeParams} the Taylor coefficients and the normalized effective mass at $\rho_0$, which characterize the $\Lambda$ potential:
\begin{align}
    &J_\Lambda=U_\Lambda(\rho_N=\rho_0,k_\Lambda=0), \label{eq:taylor1}\\
    &L_\Lambda=3\rho_N \dfrac{\partial U_\Lambda}{\partial \rho_N}\Bigr|_{\rho_N=\rho_0,k_\Lambda=0}, \label{eq:taylor2}\\
    &K_\Lambda=9\rho_N^2 \dfrac{\partial^2 U_\Lambda}{\partial \rho_N^2}\Bigr|_{\rho_N=\rho_0,k_\Lambda=0}, \label{eq:taylor3}\\
    &\dfrac{m^*_\Lambda}{m_\Lambda}\Bigr|_{\rho_N=\rho_0} = \dfrac{1}{1+\dfrac{2m_\Lambda}{\hbar^2} a^\Lambda_2 \rho_0}. \label{eq:meff}
\end{align}

\subsection{$\Lambda$ single-particle potential and $\Lambda$ binding energy}
\label{sec:LamBE}
We now present the results of the Skyrme-Hartree-Fock calculations for $\Lambda$ hypernuclei using the $\Lambda$ Skyrme interaction discussed in the previous section.

\begin{figure}[tbhp]
\includegraphics[width=\columnwidth]{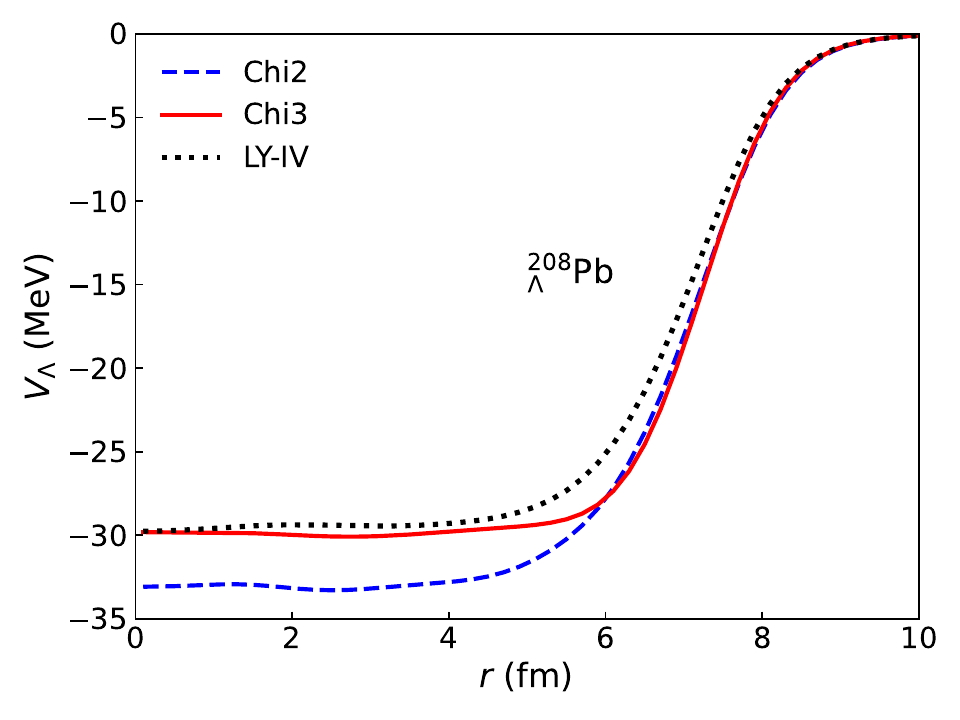}
\caption{$\Lambda$ single-particle potential~\eqref{eq:VLambda}
for hypernucleus ${}^{\rm 208}_\Lambda {\rm Pb}$ in the coordinate space. 
The dashed and solid lines show the results from the $\Lambda$ potentials Chi2 and Chi3, respectively. The dotted line corresponds to the result from the LY-IV parameter sets.
}
\label{fig:SPPot}
\end{figure}

Figure~\ref{fig:SPPot} shows the $\Lambda$ single-particle potential~\eqref{eq:VLambda} for hypernucleus ${}^{208}_\Lambda {\rm Pb}$. At a distance $r<4~\text{fm}$ where the nucleon density $\rho_N$ is close to the saturation density $\rho_0$, both Chi3 and LY-IV have the potential depth of $-30~{\text{MeV}}$ while Chi2 has a slightly greater depth of $-33~{\text{MeV}}$. Those values reflect $J_\Lambda$, the $\Lambda$-potential depth at $\rho_0$ (see Table~\ref{tab:SkyrmeParams}).

\begin{figure}[tbhp]
\includegraphics[width=\columnwidth]{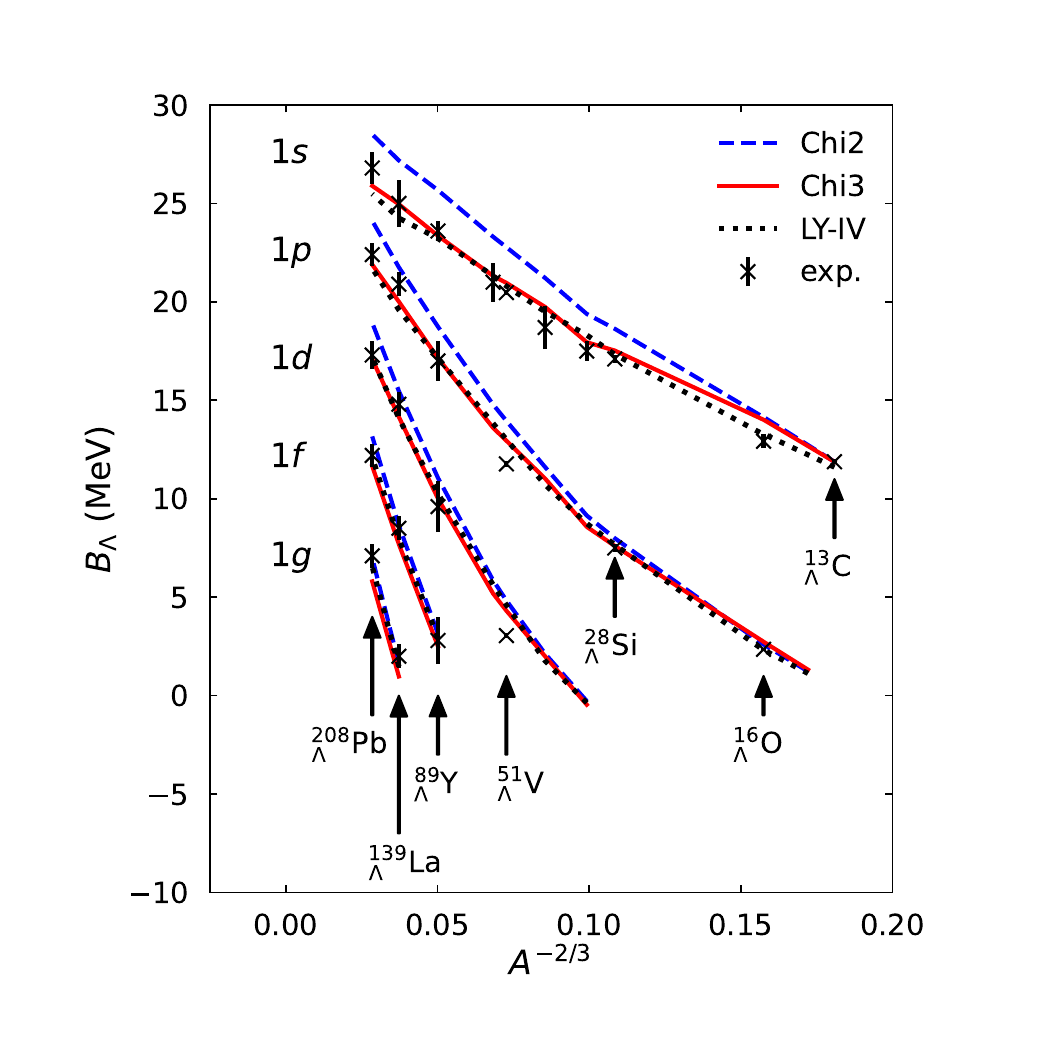}
\caption{Calculated $\Lambda$ binding energies of $1s$, $1p$, $1d$, $1f$, and $1g$ orbitals as a function of $A^{-2/3}$,
where $A$ is the mass number of hypernuclei.
Chi2 and Chi3 represent the results using the $\Lambda$ single-particle potential
with only two-body and with two- and three-body interactions from the $\chi$EFT, respectively.
The dotted line corresponds to the result using the LY-IV $\Lambda$ potential.
The crosses indicate the experimental data.
}
\label{fig:LamBEChi3}
\end{figure}

Figure~\ref{fig:LamBEChi3} compares the $\Lambda$ binding energies calculated from different $\Lambda$ potentials at mass number $A=13$--$208$
in $1s$, $1p$, $1d$, $1f$, and $1g$ orbitals. 
The experimental data at $A=16$--$208$
are listed in Table~\ref{tab:LamExp}\@.
Chi3, which includes the $\Lambda NN$ three-body force, reproduces the data.
This implies that the strong repulsive $\Lambda$ potential, which is sufficient to suppress the presence of $\Lambda$ hyperons in dense nuclear matter, is consistent with the observed $\Lambda$ hypernuclear data.
On the other hand, Chi2, which includes only the $\Lambda N$ two-body force, predicts
the overbinding of the data in the $1s$ orbital.
This is because $J_\Lambda$ is as deep as approximately $-33~{\text{MeV}}$ for Chi2.
We note that Chi2 and Chi3 have almost the same effective mass.
These results indicate the necessity of three-baryon interaction to reproduce the binding energy of $\Lambda$, which is 
consistent with the findings in Refs.~\cite{Nagels:2015lfa, Friedman:2022bpw, Friedman:2023ucs}.

The last row of Table~\ref{tab:SkyrmeParams} shows the root-mean-square deviation of the model calculation from the experimental data,
\begin{equation}
    \Delta B_\Lambda = \sqrt{ \dfrac{1}{N_\mathrm{data}}\sum_i^{N_\mathrm{data}} \left(B^\mathrm{exp}_{\Lambda,i} - B^\mathrm{cal}_{\Lambda,i}\right)^2},
    \label{eq:RMSD}
\end{equation}
where $N_\mathrm{data}$ is the number of the experimental data, and $B^{\mathrm{exp}}_{\Lambda,i}$ ($B^{\mathrm{cal}}_{\Lambda,i}$) is the experimental (calculated) $\Lambda$ binding energy of nuclide $i$. The experimental data $B^{\mathrm{exp}}_{\Lambda,i}$ are listed in Table~\ref{tab:LamExp}.
We avoid using the chi squared because it is hard to quantify the systematic error of our mean-field calculation.
Neglecting the model error in the chi-square would cause an excessive fitting to lighter hypernuclei, for which the experimental data has a smaller error while the mean-field calculation would have a larger error.
Chi3 is found to be consistent with the experimental data at the same level of accuracy as the $\Lambda$ potentials LY-IV, with the value of $\Delta B_\Lambda\approx 0.7~\text{MeV}$.
The HP$\Lambda$2 potential has a larger $\Delta B_\Lambda = 0.78~\text{MeV}$ because it is parametrized by chi-square minimization using the data without the correction of $0.5~\text{MeV}$~\cite{Gogami:2015tvu}.

\section{Parameter search for $\Lambda$ potentials}
\label{sec:ModelIndep}

In the previous section, we showed that
both
the repulsive and attractive $\Lambda$ potentials at high densities,
Chi3 and LY-IV,
reproduce the $\Lambda$ binding energy data.
This is because the $\Lambda$ binding energy is mostly determined by the potential values at low densities.
However, another attractive potential, HP$\Lambda$2, does not reproduce the data at the same level as Chi3 and LY-IV,
which suggests that the details of the potential, including the values of the Taylor coefficients are important for the $\Lambda$ binding energy.
In this section, we will investigate the parameter space of the $\Lambda$  potential that
can reproduce the binding energy of
the $\Lambda$ hypernuclear data.
From this analysis,
we expect constraints on the relations among the Taylor coefficients~\eqref{eq:taylor1}--\eqref{eq:taylor3} and the normalized effective mass~\eqref{eq:meff}.
These constraints can restrict the repulsion of the $\Lambda$ potential at high densities
similarly to the nuclear matter EOS.

\subsection{Procedure}

The symmetry-energy parameters of nuclear matter are used to examine the nuclear matter EOS at high densities using the behavior around $\rho_0$~\cite{Tews:2016jhi}. 
Similarly, to investigate the parameter space of the $\Lambda$ potential,
we parametrize the Skyrme-type $\Lambda$ potential~\eqref{eq:PotLamSkyrmeUnif} by the Taylor coefficients~\eqref{eq:taylor1}--\eqref{eq:taylor3} and the normalized effective mass $m^*_\Lambda/m_\Lambda$ at the saturation density.
We generate $13\times25\times21\times21=143325$ parameter sets of $(J_\Lambda,L_\Lambda,K_\Lambda,m^*_\Lambda/m_\Lambda)$ as combinations of the following parameter points:
\begin{subequations}
\label{eq:ParamSets}
\begin{align}
    J_\Lambda &= -33,-31.5,-32,\cdots,-27~{\text{MeV}}, \\
    L_\Lambda &= -50,-45,-40,\cdots,70~{\text{MeV}}, \\
    K_\Lambda &= 0,50,100,\cdots,1000~{\text{MeV}}, \\
    \dfrac{m^*_\Lambda}{m_\Lambda} &= 0.5, 0.525,0.55, \cdots, 1.00.
\end{align}
\end{subequations}
The range of $J_\Lambda$ is chosen to be consistent with the $\Lambda$ binding energy data of the $1s$ orbital:
$J_\Lambda$ below the range overestimates the data, and vice versa.
The effective mass $m_\Lambda^*/m_\Lambda$ has upper and lower limits because
it is sensitive to the separation of the $\Lambda$ binding energies between different orbitals.
The second derivative $K_\Lambda$ has a lower limit of $0~\rm{MeV}$
because the $\Lambda$ potential should become repulsive at high densities.
We note that these parameters in the existing models shown in Table~\ref{tab:SkyrmeParams}
fall within the above parameter ranges.

For each parameter set of $(J_\Lambda, L_\Lambda, K_\Lambda, m_\Lambda^*/m_\Lambda)$,
the Skyrme potential parameters $a^\Lambda_i$ are determined in the following procedure. First, $a^\Lambda_2$ is determined from its relation to the effective mass~\eqref{eq:meff}:
\begin{equation}
    a^\Lambda_2 = \dfrac{1}{\rho_0} \dfrac{\hbar^2}{2 m_\Lambda} \left(\dfrac{1}{m^*_\Lambda/m_\Lambda} - 1\right).
\end{equation}
The potential parameters $a^\Lambda_1$, $a^\Lambda_4$, and $a^\Lambda_5$ are determined by solving the following relations obtained from Eqs.~\eqref{eq:taylor1}--\eqref{eq:taylor3}
with $\gamma_1=1/3$ and $\gamma_2=2/3$:
\begin{subequations}
    \label{eqs:JLK}
    \begin{align}
        J_\Lambda&=a^\Lambda_1 \rho_0 + a^\Lambda_2 \tau_0 + a^\Lambda_4 \rho^{4/3}_0 + a^\Lambda_5 \rho^{5/3}_0, \\
        L_\Lambda&=3 a^\Lambda_1 \rho_0 +5 a^\Lambda_2 \tau_0 + 4 a^\Lambda_4 \rho^{4/3}_0 +5  a^\Lambda_5 \rho^{5/3}_0,\\
        K_\Lambda&=10 a^\Lambda_2 \tau_0 + 4 a^\Lambda_4 \rho^{4/3}_0 +10  a^\Lambda_5 \rho^{5/3}_0,
    \end{align}
\end{subequations}
where $\tau_0=3/5 (3 \pi^2 /2)^{2/3} \rho_0^{5/3}$.
The remaining parameter $a^\Lambda_3$ is determined to reproduce the experimental value of the $\Lambda$ binding energy of $^{13}_{\Lambda}{\rm C}$, $11.88~{\text{MeV}}$, which is taken from Ref.~\cite{Hashimoto:2006aw} with a correction of $0.5~\text{MeV}$~\cite{Gogami:2015tvu}.

Using the determined potential parameters $a_i^\Lambda$ for each parameter set, we calculate the $\Lambda$ binding energy using the Skyrme-Hartree-Fock method as explained in Sec.~\ref{sec:SHF-method}.
We evaluate the root-mean-square deviation $\Delta B_\Lambda$~\eqref{eq:RMSD}
and select the parameter sets that satisfy $\Delta B_\Lambda < 0.75~\MeV$.
We hereafter call the $\Lambda$ potentials with the selected parameter sets \textit{selected $\Lambda$ potentials}
and the others \textit{rejected $\Lambda$ potentials}.

\subsection{Results}
\label{sec:GeneratedLamPot}

\begin{figure}[tbhp]
\includegraphics[width=8cm]{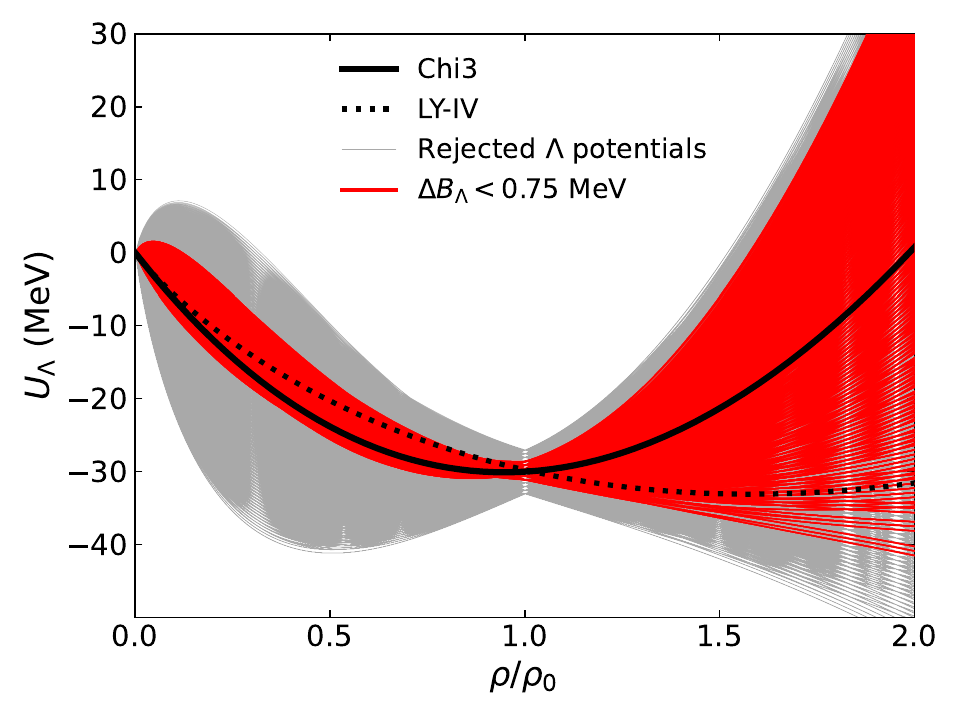}
\includegraphics[width=8cm]{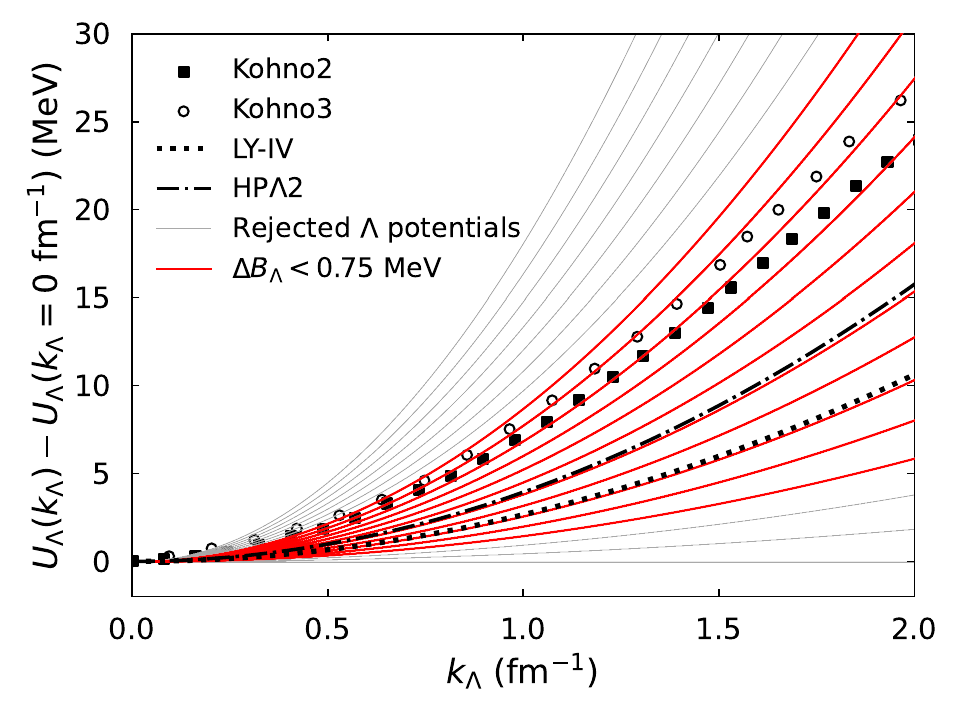}
\caption{The upper panel shows the density dependence in symmetric nuclear matter while the lower panel shows the momentum dependence of the $\Lambda$ single-particle potentials. The thin gray lines represent all generated $\Lambda$ potentials while the bold red lines are the selected $\Lambda$ potentials with $\Delta B_\Lambda<0.75~\text{MeV}$\@.
Chi2 and Chi3 represent the result using the $\Lambda$ single-particle potential
with only two-body interactions and with two- and three-body interactions from the $\chi$EFT, respectively.
The dotted line corresponds to the result using the LY-IV $\Lambda$ potential.
}
\label{fig:Pot075}
\end{figure}

We show the density dependence of the $\Lambda$ potentials
in the upper panel of Fig.~\ref{fig:Pot075}, 
and the momentum dependence of the $\Lambda$ potentials subtracting their values at $k_\Lambda=0~\text{fm}^{-1}$
in the lower panel of Fig.~\ref{fig:Pot075}.
The bold red lines indicate the selected $\Lambda$ potentials. 
We found that the density dependence at high densities $\rho>\rho_0$ spreads more widely compared to the low-density region $\rho<\rho_0$.
Namely, the $\Lambda$ binding energy data constrain the $\Lambda$ potential in the low-density region better than in the high-density region.
We note that the spread of our results at $\rho < \rho_0$ is larger than that observed in Ref.~\cite{Millener:1988hp}.
This is because we considered the finite size of the uncertainty range of the $\Lambda$ potential parameters,
while Ref.~\cite{Millener:1988hp} only shows the best fitting results by several functional forms.

In contrast to the density dependence of the $\Lambda$ potentials, the magnitude of the momentum dependence has upper and lower limits, as shown in the lower panel of Fig.~\ref{fig:Pot075}.
The LY-IV and HP$\Lambda$2 potentials lie around the lower limit while the Chi3 potentials are close to the upper limit.
The sensitivity of the energy separation between orbitals to the effective mass~\cite{Yamamoto:1988qz} constrains
the momentum dependence of the $\Lambda$ potential.

\begin{figure*}[tbhp]
\includegraphics[width=6cm]{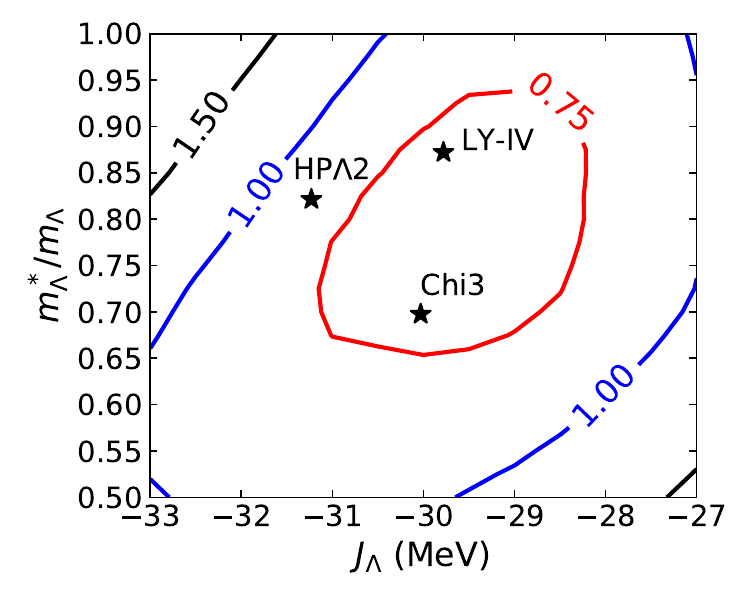}
\includegraphics[width=6cm]{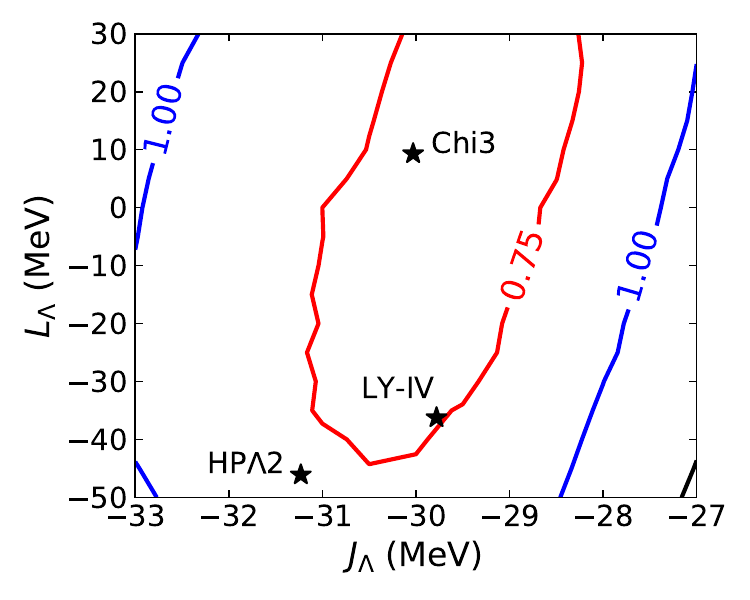}
\includegraphics[width=6cm]{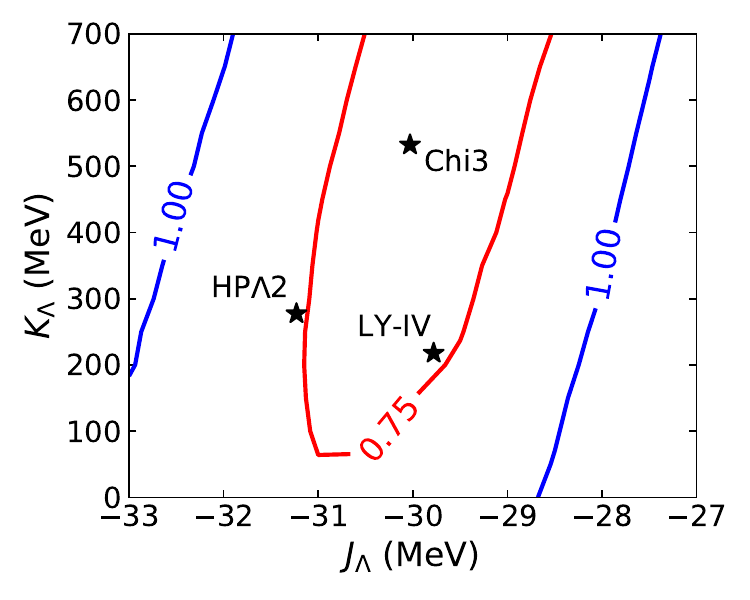}
\includegraphics[width=6cm]{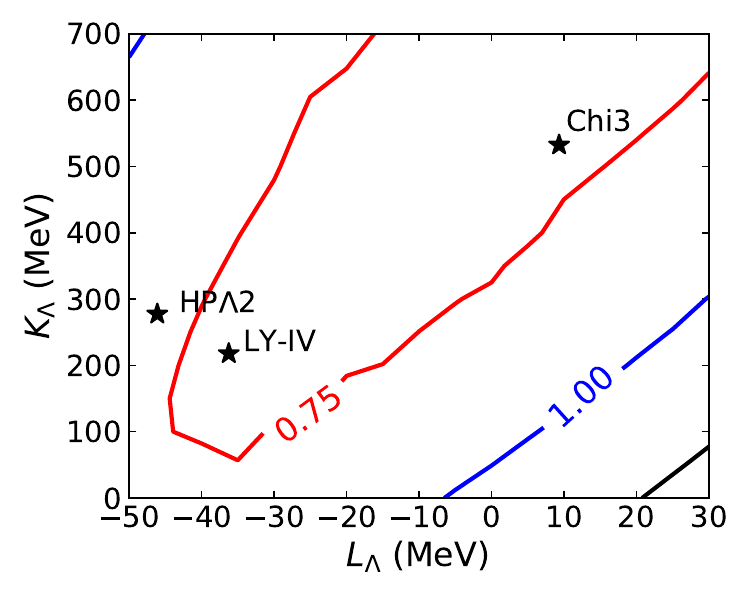}
\caption{Contour plot of the root-mean-square deviation $\Delta B_\Lambda$ of the calculated $\Lambda$ binding energy from the experimental data.
The stars indicate the points corresponding to the two parameters of Chi3, LY-IV, and HP$\Lambda$2.
Note that the other parameters of those specific models do not necessarily match the optimized ones.
Top left panel: $\Delta B_\Lambda$ as a function of the depth of the $\Lambda$ potential at the saturation density, $J_\Lambda$, and the normalized effective mass $m^*_\Lambda/m_\Lambda$.
The first and second derivatives, $L_\Lambda$ and $K_\Lambda$, are optimized for minimizing $\Delta B_\Lambda$ for each point of $(J_\Lambda, m^*_\Lambda/m_\Lambda)$ by using the golden-section search.
Top right panel: $\Delta B_\Lambda$ as a function of
the depth of the $\Lambda$ potential, $J_\Lambda$, and the first derivative $L_\Lambda$,
where $(K_\Lambda,m^*_\Lambda/m_\Lambda)$ are optimized.
Bottom left panel: $\Delta B_\Lambda$ as a function of
the depth of $\Lambda$ potential $J_\Lambda$ and the second derivative $K_\Lambda,$
where $(L_\Lambda,m^*_\Lambda/m_\Lambda)$ are optimized.
Bottom right panel: $\Delta B_\Lambda$ as a function of
the first derivative $L_\Lambda$ and the second derivative $K_\Lambda$,
where $(J_\Lambda,m^*_\Lambda/m_\Lambda)$ are optimized.
}
\label{fig:Jm_best}
\end{figure*}

Let us now examine the correlations among the parameters of the selected $\Lambda$ potentials.
For this purpose,
we choose two parameters from $(J_\Lambda, L_\Lambda, K_\Lambda, m^*_\Lambda/m_\Lambda)$ and plot $\Delta B_\Lambda$ as a function of the chosen parameters. 
We consider several choices of the two parameters as shown in Fig.~\ref{fig:Jm_best}.
In calculating $\Delta B_\Lambda$, the other parameters are optimized to minimize $\Delta B_\Lambda$
by using the golden-section search~\cite{Kiefer:1953GSS} within the ranges in Eqs.~\eqref{eq:ParamSets}.
For example, for the $J_\Lambda$-$m^*_\Lambda/m_\Lambda$ plot in the top left panel of Fig.~\ref{fig:Jm_best}, the remaining parameters, $L_\Lambda$ and $K_\Lambda$, are optimized for each point of $(J_\Lambda, m^*_\Lambda/m_\Lambda)$.
The values of the optimized parameters are shown in Appendix~\ref{app:OptParams}.

\begin{figure*}[tbhp]
\includegraphics[width=5.9cm]{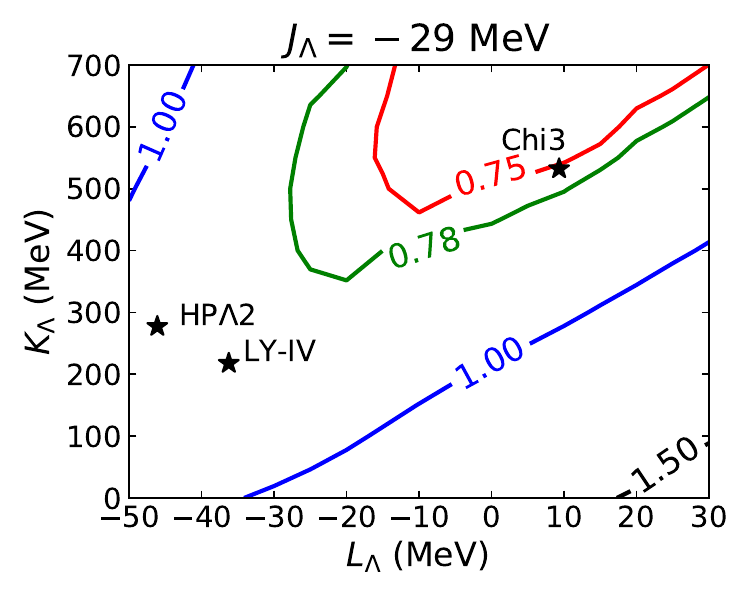}
\includegraphics[width=5.9cm]{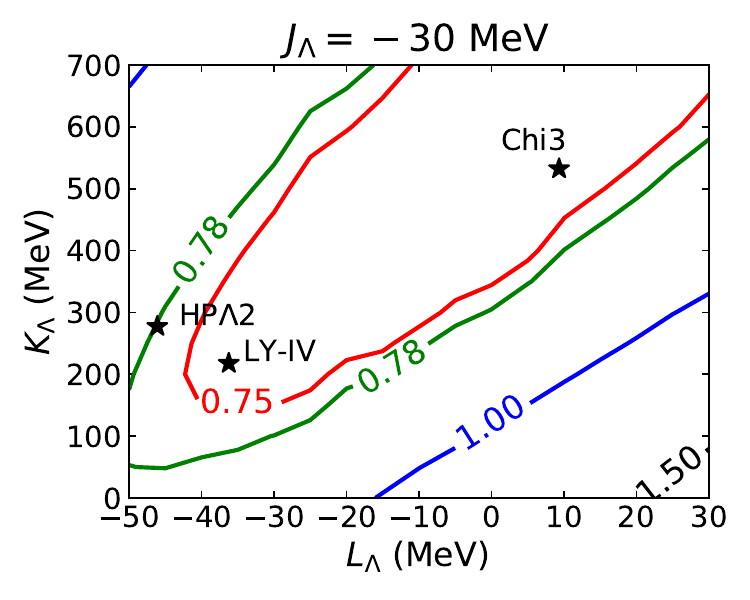}
\includegraphics[width=5.9cm]{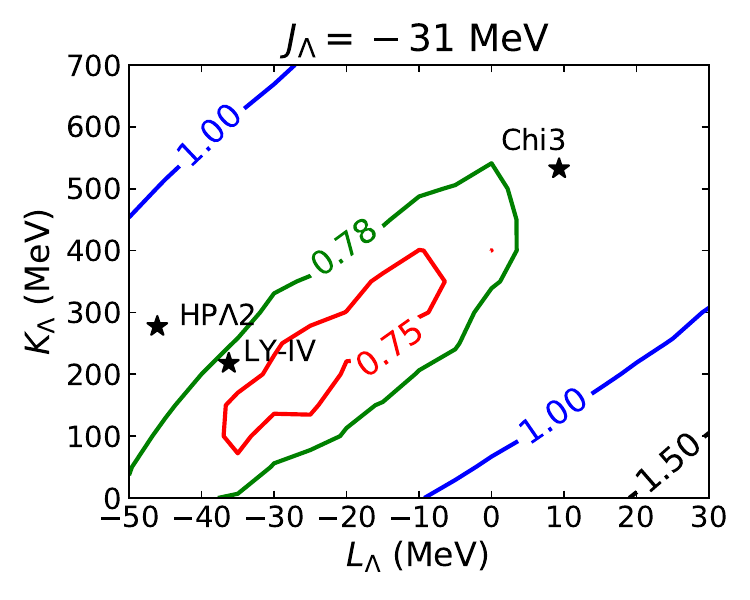}
\caption{Same as the bottom right panel of Fig.~\ref{fig:Jm_best}, but for three different fixed values of $J_\Lambda$.
The normalized effective mass $m^*_\Lambda/m_\Lambda$ is optimized for minimizing $\Delta B_\Lambda$ for each $(J_\Lambda, L_\Lambda, K_\Lambda)$ by using the golden-section search.
}
\label{fig:LK_J}
\end{figure*}
The top left panel of Fig.~\ref{fig:Jm_best}
shows $\Delta B_\Lambda$ as functions of $J_\Lambda$ and $m^*_\Lambda/m_\Lambda$.
The selected parameter sets are 
within the ranges of $-31.5 < J_\Lambda < -28~{\text{MeV}}$
and $0.65 < m^*_\Lambda/m_\Lambda < 0.95$.
The potential depth $J_\Lambda$ is consistent with the well-known results by the Woods-Saxon potential~\cite{Millener:1988hp,Gal2016}.
A positive correlation between $J_\Lambda$ and $m^*_\Lambda/m_\Lambda$ is found because the contribution of the kinetic-energy term $-\hbar^2\nabla_\Lambda^2/2m^*_\Lambda$ increases
as the normalized effective mass $m^*_\Lambda/m_\Lambda$ decreases.
The HP$\Lambda$2 potential is located outside of the region of the selected $\Lambda$ potentials because HP$\Lambda$2 was originally constructed through the chi-square minimization using the data without the $0.5~\text{MeV}$ correction~\cite{Gogami:2015tvu}.
In the top right panel of Fig.~\ref{fig:Jm_best}, we show $\Delta B_\Lambda$ as functions of  $J_\Lambda$ and $L_\Lambda$. 
The region of the selected potentials
is bounded from below,
$L_\Lambda \ge -45~\text{MeV}$
because the experimental data of the $\Lambda$ binding energy cannot be reproduced with the $\Lambda$ potential that is too shallow at $\rho<\rho_0$.
In the bottom left panel of Fig.~\ref{fig:Jm_best}, we show $\Delta B_\Lambda$ as functions of $J_\Lambda$ and $K_\Lambda$.
The condition $\Delta B_\Lambda <0.75~\text{MeV}$ is satisfied for
$K_\Lambda \ge 50~\text{MeV}$.
There is no upper limit in $K_\Lambda$,
which implies that the $\Lambda$ potential can be even more repulsive than the ones covered in this parameter search.
In the bottom right panel of Fig.~\ref{fig:Jm_best}, we show $\Delta B_\Lambda$ as functions of  $L_\Lambda$ and $K_\Lambda$.
For the region with $\Delta B_\Lambda<0.75~{\text{MeV}}$, there is a positive correlation between $L_\Lambda$ and $K_\Lambda$
so that the effects of the two parameters compensate for each other at low densities
at which the $\Lambda$ potential is constrained.
Larger values of $L_\Lambda$ make the potential deeper at $\rho<\rho_0$ while larger values of $K_\Lambda$ make it shallower.
Nevertheless, the uncertainty region with $\Delta B_\Lambda<0.75~\text{MeV}$ is not small enough, i.e., the sizes of the region are about $40$ and $400~\text{MeV}$ for $L_\Lambda$ and $K_\Lambda$, respectively. This reflects the fact that the $\Lambda$ potential at $\rho<\rho_0$ is not sufficiently limited to discriminate Chi3 from LY-IV.

It should be noted that the $\Delta B_\Lambda$ values of the specific models (e.g., LY-IV) in Table.~\ref{tab:SkyrmeParams} do not necessarily match $\Delta B_\Lambda$ at their locations in Figs.~\ref{fig:Jm_best} and~\ref{fig:LK_J}.
This is because the other parameters (i.e., optimized parameters and $a_3^\Lambda$) are different from the parameters in the specific model.
The $\Delta B_\Lambda$ value depends on $(J_\Lambda, L_\Lambda, K_\Lambda,  m^*_\Lambda/m_\Lambda, a_3^\Lambda)$, but only two of them are the same with the specific models in each panel of Figs.~\ref{fig:Jm_best} and~\ref{fig:LK_J}.
For example, in the top left panel of Fig.~\ref{fig:Jm_best}, the other parameters $(K_\Lambda, L_\Lambda, a_3^\Lambda)$ are determined for each pair of $(J_\Lambda, m^*_\Lambda/m_\Lambda)$ so that they minimize $\Delta B_\Lambda$ while keeping the $\Lambda$ binding energy of $^{13}_\Lambda\mathrm{C}$, $B_\Lambda$($^{13}_\Lambda\mathrm{C}$).
However, such a set of parameters does not necessarily match the parameters of the specific models,
which are determined by different criteria. 
Specifically, we use the up-to-date value of $B_\Lambda(^{13}_\Lambda\mathrm{C})=11.88~\text{MeV}$ in our work to 
fix the value of $a^\Lambda_3$, while $B_\Lambda(^{13}_\Lambda\mathrm{C})=11.69~\text{MeV}$ is used in LY-IV~\cite{Lanskoy:1997xq}.
It should also be noted that the $\Delta B_\Lambda$ value at the location of a specific model (e.g., LY-IV) differs for each panel because the optimized parameter sets at the LY-IV locations are different for different panels. Only two parameters are matched to LY-IV in each panel, and the combination of the two parameters differs for each panel.

We show $\Delta B_\Lambda$ for three different values of $J_\Lambda$ as functions of $L_\Lambda$ and $K_\Lambda$ in Fig.~\ref{fig:LK_J}, where the parameter $m^*_\Lambda/m_\Lambda$ is optimized to minimize $\Delta B_\Lambda$ for each $(J_\Lambda, L_\Lambda, K_\Lambda)$
employing the golden-section search.
For $J_\Lambda=-29~\text{MeV}$,
we see that the parameter region of the $\Lambda$ potentials with small $\Delta B_\Lambda$ have larger $L_\Lambda$ and $K_\Lambda$ compared to the case of unconstrained $J_\Lambda$ (the bottom right panel of Fig.~\ref{fig:Jm_best}).
This is because to explain the $\Lambda$ binding energy with a shallow potential, the $\Lambda$ potential at $\rho\lesssim\rho_0$ has to be deeper by taking large $L_\Lambda$.
For $J_\Lambda=-31~\text{MeV}$, the parameter region of $\Delta B_\Lambda<0.78~\text{MeV}$ has upper limits at $L_\Lambda=0~\text{MeV}$ and $K_\Lambda=550~\text{MeV}$.
In contrast to the case of $J_\Lambda=-29~\text{MeV}$, the $\Lambda$ potential at $\rho\lesssim \rho_0$ has to be shallower by taking small $L_\Lambda$.
Thus, measuring $J_\Lambda$ within the error of $1~\text{MeV}$ enables us to constrain the range of $L_\Lambda$ and then $K_\Lambda$ through the positive correlation.
This helps to discuss whether the $\Lambda$ potential is repulsive or attractive at high densities.
For $J_\Lambda=-30~\text{MeV}$,
the parameter region within $\Delta B_\Lambda<0.75~\text{MeV}$
includes various $K_\Lambda$ values that are the same as those in the bottom right panel of Fig.~\ref{fig:Jm_best}.
Therefore, if $J_\Lambda$ is determined as $-30~\text{MeV}$ by high-resolution data of the $\Lambda$ binding energy,
further constraint at $\rho<\rho_0$ is needed to determine $L_\Lambda$ and $K_\Lambda$.

\section{$\Lambda$ admixture in neutron star matter}
\label{sec:LamAdmix}
In the previous section, we examined the parameter space of $\Lambda$ potentials consistent with the $\Lambda$ binding energy data.
In this section, we discuss the admixture of $\Lambda$ in neutron stars by the chemical potentials and investigate the parameter region where $\Lambda$'s do not appear in neutron stars.
The $\Lambda$ hyperons appear in neutron star matter at the baryon density $\rho$ when
the minimal chemical potential of $\Lambda$,
\begin{equation}
    \mu^0_\Lambda(\rho) = m_\Lambda c^2 + U_\Lambda(\rho,k_\Lambda=0),
\end{equation}
exceeds the neutron chemical potential.

We determine the neutron chemical potential in neutron star ($npe\mu$) matter
by solving the $\beta$-equilibrium conditions,
\begin{align}
    \label{eq:npe}
    \mu_n &= \mu_p + \mu_e, \\
    \label{eq:emu}
    \mu_e &= \mu_\mu,
\end{align}
together with the baryon number conservation and the charge neutrality condition,
\begin{align}
    \rho &= \rho_p + \rho_n, \\
    \rho_p &= \rho_e + \rho_\mu,
\end{align}
where
\begin{equation}
    \mu_i = \biggl(\dfrac{\partial \tilde{\mathcal{E}}}{\partial \rho_i}\biggr)_{\rho_{j}\neq \rho_i}, \quad i,j=n,p,e,\mu,
\end{equation}
are the chemical potentials of the matter constituents, and
$\rho_i$ ($i = n, p, e, \mu$) are the corresponding densities.
The total energy density $\tilde{\mathcal{E}}$ is given as
\begin{align}
    \tilde{\mathcal{E}}(\rho_n,\rho_p,\rho_e,\rho_\mu) =& \tilde{\mathcal{E}}_N(\rho_n, \rho_p) + m_n c^2 \rho_n + m_p c^2 \rho_p \nonumber \\
    &+ \tilde{\mathcal{E}}_e(\rho_e) + \tilde{\mathcal{E}_\mu}(\rho_\mu),
\end{align}
where $\tilde{\mathcal{E}}_N$ is the nucleon energy density, and 
$c$ is the speed of light.
The energy densities of electrons $\tilde{\mathcal{E}}_e$ and muons $\tilde{\mathcal{E}}_\mu$ are assumed to be those of the Fermi gas.
For the energy density of nucleons $\tilde{\mathcal{E}}_N$, we use the SLy4~\cite{Chabanat:1997un} and BSk24~\cite{Goriely:2013xba} parameter sets.
Both are in good agreement with the pure neutron matter EOS from the $\chi$EFT up to the next-to-next-to-next-to-leading order (N3LO)~\cite{Drischler:2020hwi,Burgio:2021vgk}.
SLy4 is a softer EOS with the maximum neutron star mass of 2.06$M_\odot$ compared
to BSk24 with 2.28$M_\odot$.

\begin{figure}[tbhp]
\includegraphics[width=8cm]{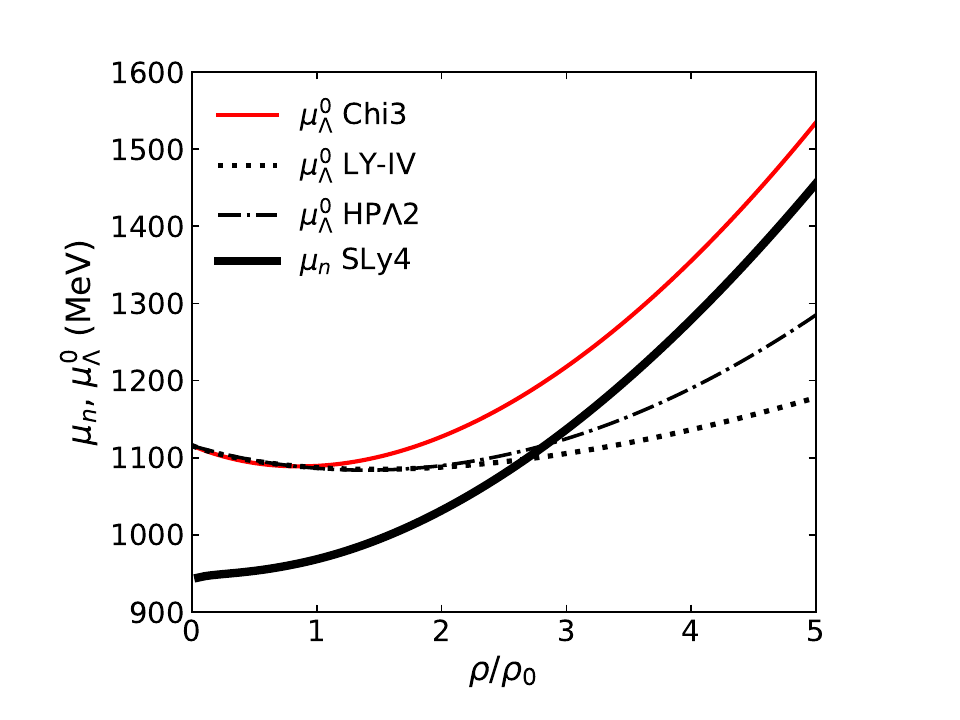}
\includegraphics[width=8cm]{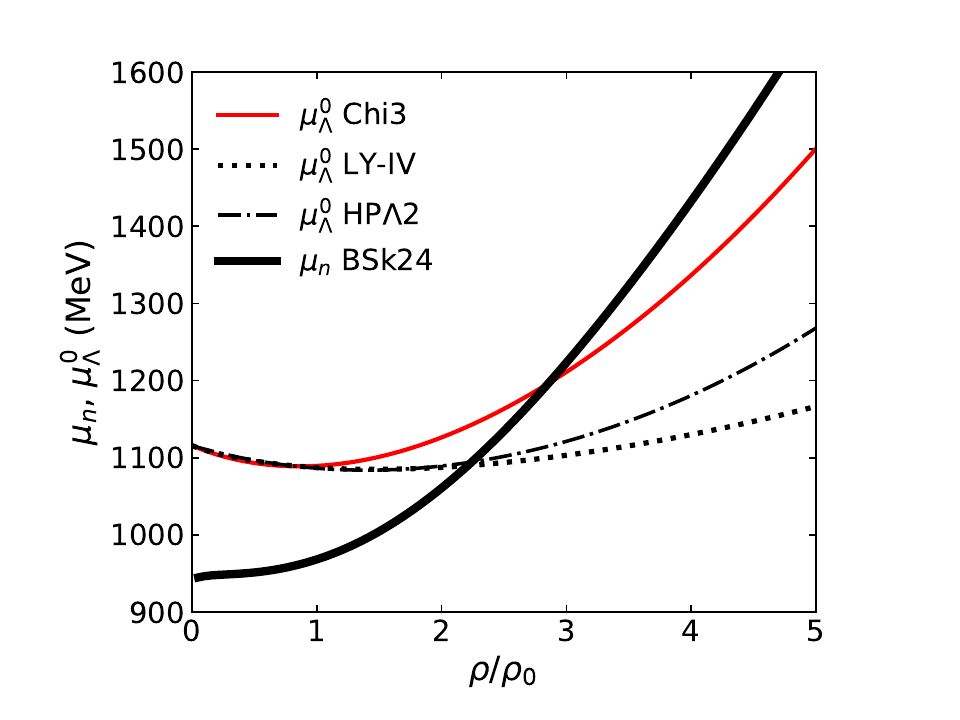}
\caption{$\Lambda$ chemical potentials at zero momentum and neutron chemical potential in neutron star ($npe\mu$) matter are depicted as a function of the normalized baryon density.
The solid, dotted, and dash-dotted lines correspond to the $\Lambda$ chemical potential at zero momentum for Chi3, LY-IV, and HP$\Lambda$2, respectively.
The bold solid line represents the neutron chemical potential calculated by using the SLy4 (upper panel) and BSk24 (lower panel) parameter sets.
}
\label{fig:ChemicalPot}
\end{figure}

In Fig.~\ref{fig:ChemicalPot}, we compare the minimal chemical potential of $\Lambda$, $\mu^0_\Lambda$, and $\mu_n$ in the neutron star matter as a function of the normalized baryon density $\rho/\rho_0$.
In the top panel, we confirm that the minimal $\Lambda$ chemical potential of Chi3 is larger than the neutron chemical potential at $\rho/\rho_0\leq5$
as reported in Ref.~\cite{Gerstung:2020ktv}.
On the other hand, by using the $\Lambda$ potentials of LY-IV and HP$\Lambda$2, $\Lambda$ hyperons are found to admix in neutron stars in the density range $2$--$3\rho_0$, as found in phenomenological models with hyperons~\cite{Glendenning:1991es, Schaffner:1995th,Balberg:1997yw,Baldo:1999rq, Vidana:2000ew} causing the softening of the EOS.
In the bottom panel, all the $\Lambda$-potential models are found to exhibit the appearance of $\Lambda$'s in neutron stars,
representing that the appearance of $\Lambda$'s depends on the model of the nucleonic EOS.

\begin{figure}[tbhp]
\includegraphics[width=8cm]{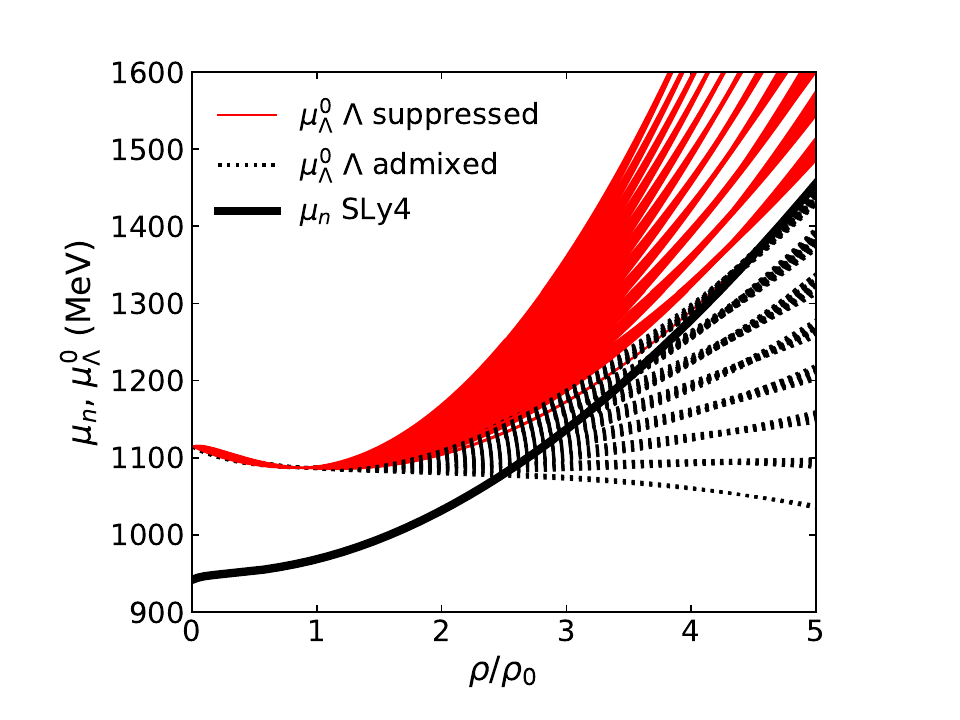}
\includegraphics[width=8cm]{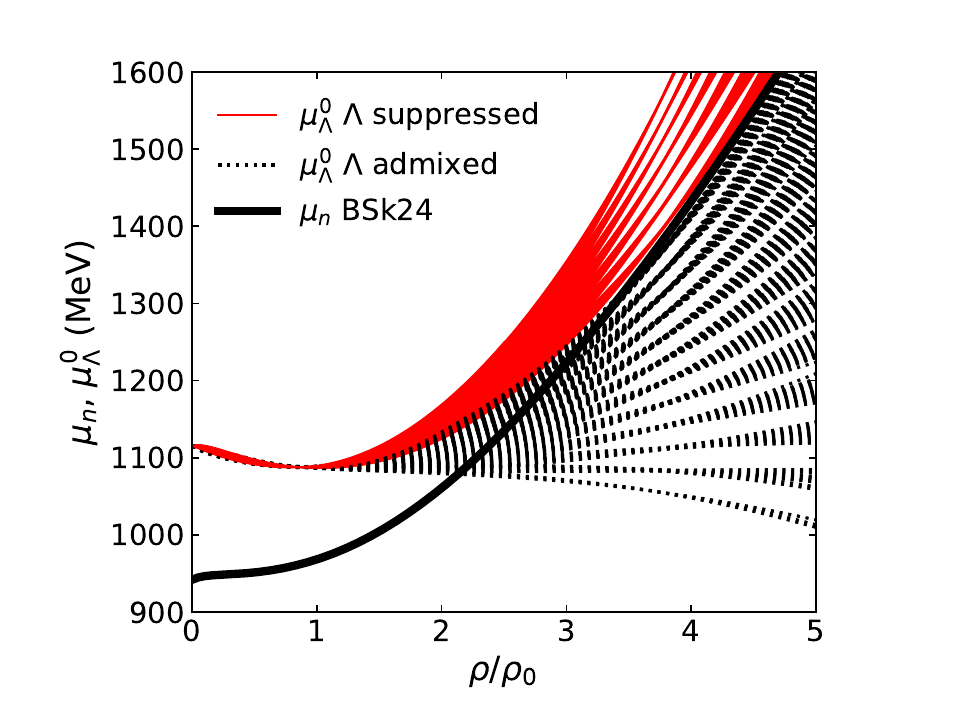}
\caption{Same as Fig.~\ref{fig:ChemicalPot},
but the $\Lambda$ chemical potentials at zero momentum using the density dependence of the selected
$\Lambda$ potentials in Fig.~\ref{fig:Pot075} are shown.
The solid red lines are the $\Lambda$ potentials with $\mu_\Lambda^0>\mu_n$ at $\rho/\rho_0\leq5$, which suppress the $\Lambda$ hyperons in neutron star matter,
while the dotted lines are those which fulfill the condition $\mu^0_\Lambda\leq\mu_n$ at some densities so that $\Lambda$ hyperons are admixed.
}
\label{fig:ChemicalPotAdmix}
\end{figure}

In Fig.~\ref{fig:ChemicalPotAdmix}, the minimal $\Lambda$ chemical potentials calculated from the selected $\Lambda$ potentials shown in Fig.~\ref{fig:Pot075} are compared with the neutron chemical potential in neutron star matter. 
The solid red lines correspond to the minimum $\Lambda$ chemical potentials that suppress the appearance of $\Lambda$ in neutron matter, i.e.,
$\mu^0_\Lambda >\mu_n$ at $\rho/\rho_0\leq5$.
We confirmed that
the suppression of $\Lambda$ appearance
hardly changes even when we check it up to the central density of a maximum-mass neutron star ($\rho \leq \rho_\mathrm{c}^\mathrm{max}$) instead of $\rho/\rho_0 \leq 5$: the number of $\Lambda$ potentials with no $\Lambda$ appearance decreases only by two for SLy4 with $\rho_\mathrm{c}^\mathrm{max}=7.5\rho_0$ and is unchanged for BSk24 with $\rho_\mathrm{c}^\mathrm{max}=9.4\rho_0$,
where $\rho_\mathrm{c}^\mathrm{max}$ is taken from the CompOSE database~\cite{Typel:2013rza,*Oertel:2016bki,*CompOSECoreTeam:2022ddl} for each EOS.
For $\rho>3.5\rho_0$, the minimum $\Lambda$ chemical potentials are split into 19 groups
corresponding to
$K_\Lambda=100$, $150$, $200$, $\cdots$, and $1000~\text{MeV}$.
We note that the gaps between the 19 groups would be filled by considering more points in $K_\Lambda$.
We also note that several parameter sets with different $L_\Lambda$'s and $J_\Lambda$'s are degenerate in each group, which means that
the high-density part of the $\Lambda$ chemical potential is mostly determined by $K_\Lambda$.
We argue that the onset of $\Lambda$ hyperons in neutron stars can be judged using the value of $K_\Lambda$:
with the SLy4 EOS $\Lambda$'s do not appear in neutron stars if $K_\Lambda\geq500~\text{MeV}$, while with the BSk24 EOS $\Lambda$'s do not appear if $K_\Lambda\geq700~\text{MeV}$.
Therefore, the second derivative of the $\Lambda$ potential, $K_\Lambda$, would be the important parameter in discussing the admixture of $\Lambda$'s in neutron star matter.

\section{Summary}
\label{sec:summary}
We have examined
the $\Lambda$ potentials using
the binding energies of $\Lambda$ in hypernuclei within the Skyrme-Hartree-Fock method with spherical symmetry.
It is found that the $\Lambda$ potential from the $\chi$EFT~\cite{Gerstung:2020ktv, Kohno:2018gby}
with two- and three-body force
reproduces the experimental $\Lambda$ binding energy data at the same level of accuracy as 
LY-IV~\cite{Lanskoy:1997xq} potential
while $\chi$EFT with only two-body force overestimates the $\Lambda$ binding energy.
Thus, the $\chi$EFT $\Lambda$ potential which suppresses $\Lambda$ in dense neutron star matter is consistent with the
$\Lambda$ binding energy.
Taken together with our previous work~\cite{Nara:2022kbb}, we conclude that the $\Lambda$-suppressed scenario is consistent with the $\Lambda$ directed flow data of heavy-ion collisions and the $\Lambda$ binding energy data of hypernuclei.
More detailed studies are necessary in future work considering the $YY$ and $YYN$ interactions.

Next, we search for the parameter space of the $\Lambda$ potentials by varying the Taylor coefficients and the effective mass at the saturation density.
The root-mean-square deviation $\Delta B_\Lambda$ is used to evaluate the consistency between the calculated $\Lambda$ binding energies and experimental data.
It is shown that the depth of the $\Lambda$ potential $J_\Lambda$ is constrained from the $\Lambda$ binding energy data within the accuracy of 
$-31.5 < J_\Lambda < -28~{\text{MeV}}$.
These values are consistent with the well-known Woods-Saxon results~\cite{Gal2016}. 
There are lower and upper limits for the normalized effective mass: $0.65 < m^*_\Lambda/m_\Lambda < 0.95$. This reflects the fact that the energy splitting between orbitals is sensitive to the effective mass.

A positive correlation between the first- and second-order Taylor coefficients of the $\Lambda$ potential, $L_\Lambda$ and $K_\Lambda$,
is found, which reflects the fact that the $\Lambda$ potential at $\rho<\rho_0$ is constrained.
It is shown that $K_\Lambda$ can be well constrained by determining $J_\Lambda$
within the accuracy of $1~\text{MeV}$, i.e.,
$K_\Lambda>350~\text{MeV}$ is favored for $J_\Lambda=-29~\text{MeV}$, while $K_\Lambda<550~\text{MeV}$ is favored for $J_\Lambda=-31~\text{MeV}$.
In the future, the value of $J_\Lambda$ is expected to be determined more precisely through high-resolution data obtained at the Japan Proton Accelerator Research Complex (J-PARC)~\cite{Aoki:2021cqa}.
These data can be used to constrain 
the second-order Taylor coefficient $K_\Lambda$.
Also, the $\Lambda$ potentials with $K_\Lambda\geq500~{\text{MeV}}$
are found to suppress $\Lambda$s in $\beta$-stable neutron star matter for SLy4,
while the $\Lambda$ potentials with $K_\Lambda\geq700~{\text{MeV}}$ for BSk24 suppress $\Lambda$.
Therefore,
the determination of $J_\Lambda$ helps discriminate the $\Lambda$-avoiding scenario from the $\Lambda$-admixing scenario in neutron stars.

To precisely determine the $\Lambda$ potential at high densities, it would also be important to investigate other experimental data in future works.
For example, the observables in heavy-ion collisions
may be sensitive to the $\Lambda$ potential at high densities:
the elliptic flow including the centrality dependence
and nuclear cluster production can be affected by the $\Lambda$ potential.
Also, the hypertriton ${}^3_\Lambda {\rm H}$ directed flow, which was recently measured by the STAR Collaboration~\cite{STAR:2022fnj},
would be a promising future work.
For another example,
the value of $K_\Lambda$ could be constrained by the excitation spectra of the breathing mode for $\Lambda$ hypernuclei~\cite{HirokazuTamura2023} in the same way as the incompressibility of  nuclear matter, $K$.
The $\Lambda$ hypernuclei with large surface, such as neutron-rich hypernuclei, might also give constraints on the $\Lambda$ potential.

\begin{acknowledgments}
We thank Prof. Hirokazu Tamura for his useful comments and Prof. Kouichi Hagino for his careful reading of the manuscript and helpful comments.
This work was supported in part by the
Grants-in-Aid for Scientific Research from JSPS
(Grants No. JP21K03577, 
No. JP19H01898, 
No. JP21H00121, 
and No. JP23K13102).
This work was also supported by JST,
the establishment of university fellowships towards
the creation of science technology innovation,
Grant No. JPMJFS2123.
\end{acknowledgments}

\appendix

\section{Explicit expressions of the Skyrme-Hartree-Fock potentials and densities}
\label{app:expressions}

\begin{table}[tbhp]
\caption{SLy4~\cite{Chabanat:1997un} parameter set}
\centering
\begin{tabular}{l@{\qquad}r}
\hline
\hline
Parameter & Value \\
\hline
$t_0~({{\text{MeV}}~\text{fm}^3})$ & $-2488.91$ \\
$t_1~({{\text{MeV}}~\text{fm}^5})$ & $486.82$ \\
$t_2~({{\text{MeV}}~\text{fm}^5})$ & $-546.39$ \\
$t_3~({{\text{MeV}}~\text{fm}}^{3+3\gamma})$ & $13777.0$ \\
$x_0$ & $0.834$ \\
$x_1$ & $-0.344$ \\
$x_2$ & $-1.000$ \\
$x_3$ & $1.354$ \\
$\gamma$ & $1/6$ \\
$W_0~({{\text{MeV}}~\text{fm}^5})$ & $123.0$ \\
\hline
\hline
\end{tabular}
\label{table:SLy4}
\end{table}

Here, explicit forms of the terms appearing in the Skyrme-Hartree-Fock equation~\eqref{eq:SHFeq}
are given as in Refs.~~\cite{Chabanat:1997un,Guleria:2011kk}.
The effective mass is defined by
\begin{align}
    \dfrac{\hbar^2}{2m^*_q} =& \dfrac{\hbar^2}{2m_q} 
    + \dfrac{1}{8} \left[ t_1 (2 + x_1)  t_2(2+x_2) \right] \rho_N \nonumber \\
    &+ \dfrac{1}{8} \left[ t_1 (2 x_1 + 1)  - t_1 (2 x_1 + 1) \right] \rho_q +a^\Lambda_2 \rho_\Lambda.
\end{align}
The single-particle and spin-orbit potentials are given by
\begin{align}
    V_q =& V^N_q + V^\Lambda_q + \delta_{q,p} V_{\mathrm{coul}}, \\
    W_q =& \dfrac{1}{2} W_0 \dfrac{\mathrm{d}}{\mathrm{d}r}\left[\rho_N+\rho_q\right] -\dfrac{1}{8} \left( t_1 x_1  + t_2 x_2 \right) J_N(r) \nonumber \\
    &+ \dfrac{1}{8} (t_1 - t_2) J_q(r),
\end{align}
respectively, where
\begin{align}
    V^N_q =& \dfrac{1}{2} t_0 (2+x_0) \rho_N + \dfrac{1}{2} t_0 (2 x_0 + 1) \rho_q \nonumber \\
    &+ (\gamma+2) \dfrac{1}{24} t_3 (2+x_3) \rho^{\gamma+1}_N \nonumber \\
    &+ \dfrac{1}{24} t_3 (2 x_3+1) \left[\gamma \rho^{\gamma-1}_N (\rho^2_p + \rho^2_n) + 2\rho^\gamma_N \rho_q\right] \nonumber \\
    &+ \dfrac{1}{8} \left( t_1 (2 + x_1)  t_2(2+x_2) \right) \tau_N \nonumber \\
    &+ \dfrac{1}{8} \left( t_1 (2 x_1 + 1)  - t_1 (2 x_1 + 1) \right) \tau_q \nonumber \\
    &- \dfrac{1}{16} \left( 3 t_1 (2 + x_1)  - t_2 (2 + x_2) \right) \Delta\rho_N \nonumber \\
    &- \dfrac{1}{16} \left( 3 t_1 (2 x_1 + 1)  + t_2 (2 x_2 + 1) \right) \Delta \rho_q \nonumber \\
    &- \dfrac{1}{2}W_0 \left[\nabla \cdot \bm{J}_N  + \nabla \cdot \bm{J}_q \right], \\
    V^\Lambda_q =& a^\Lambda_1 \rho_\Lambda + a^\Lambda_2 \tau_\Lambda - a^\Lambda_3 \Delta{\rho_\Lambda} \nonumber \\
    & + (1+\gamma_1) a^\Lambda_4 {\rho_N}^{\gamma_1} \rho_\Lambda + (1+\gamma_2) a^\Lambda_5 {\rho_N}^{\gamma_2} \rho_\Lambda, \\
    V_{\mathrm{coul}} =& e^2 \int \mathrm{d}^3{r'} \dfrac{\rho_p ({\bm{r'}})}{\left|\bm{r}-\bm{r'}\right|} - e^2 \left(\dfrac{3\rho_p}{\pi}\right)^{1/3}.
\end{align}
The values of the parameters $t_i$, $x_i$, $\gamma$, and $W_0$ are given by the SLy4 parameter set shown in Table~\ref{table:SLy4}.
The isospins $q=\pm1/2$ specify the neutron and proton, respectively.
The density $\rho_q$ and the kinetic density $\tau_q$ are given by Eq.~\eqref{eq:Ndensity} while
the spin density $\bm{J}_q$ is written as

\begin{align}
    \bm{J}_q(\bm{r}) &= -i \sum_i \phi^*_i(\bm{r},q) \nabla \phi_i(\bm{r},q) \times \langle\sigma' | \bm{\sigma} | \sigma \rangle.
\end{align}
The nucleon density, the kinetic density, and the spin density are defined as $\rho_N=\sum_q \rho_q$, $\tau_N=\sum_q \tau_q$, and $\bm{J}_N=\sum_q \bm{J}_q$, respectively.

\section{$\Lambda$ binding energy data}
The $\Lambda$ binding energy data used in this study
are summarized in Table~\ref{tab:LamExp}.
The data measured in the $(\pi^+, K^+)$ experiments listed in Ref.~\cite{Hashimoto:2006aw} (${}^{16}_\Lambda{\rm O}, {}^{28}_\Lambda{\rm Si}, {}^{51}_\Lambda{\rm V}, {}^{139}_\Lambda{\rm La}, {}^{208}_\Lambda{\rm Pb}$) are used with the modification $0.5~{\text{MeV}}$ as pointed out in Ref.~\cite{Gogami:2015tvu}.

\begin{table*}[tbhp]
\caption{Experimental data of $\Lambda$ binding energy (B.E.) for various hypernuclei used in this work.}
\begin{tabular}{llllll}
\hline
\hline
Hypernuclei & B.E. (${\text{MeV}}$) & & & & \\ \cline{2-6}
 & $1s$ & $1p$ & $1d$ & $1f$ & $1g$ \\
\hline
${}^{16}_\Lambda{\rm O}$~\cite{Hashimoto:2006aw} & $12.92\pm0.35$ & $2.35\pm0.05$ & & & \\
${}^{28}_\Lambda{\rm Si}$~\cite{Hashimoto:2006aw} & $17.1\pm0.2$ & $7.5\pm0.2$ & & & \\
${}^{32}_\Lambda{\rm S}$~\cite{Bando:1990yi} & $17.5\pm0.5$ & & & & \\
${}^{40}_\Lambda{\rm Ca}$~\cite{Pile:1991cf} & $18.7\pm1.1$ & & & & \\
${}^{51}_\Lambda{\rm V}$~\cite{Hotchi:2001rx,Hashimoto:2006aw} & $20.47\pm0.13$ & $11.77\pm0.16$ & $3.05\pm0.13$ & & \\
${}^{56}_\Lambda{\rm Fe}$~\cite{Bando:1990yi} & $21.0\pm1.0$ & & & & \\
${}^{89}_\Lambda{\rm Y}$~\cite{Hashimoto:2006aw} & $23.6\pm0.5$ & $17.0\pm1.0$ & $9.6\pm1.3$ & $2.8\pm1.2$ & \\
${}^{139}_\Lambda{\rm La}$~\cite{Hashimoto:2006aw} & $25.0\pm1.2$ & $20.9\pm0.6$ & $14.8\pm0.6$ & $8.5\pm0.6$ & $2.0\pm0.6$ \\
${}^{208}_\Lambda{\rm Pb}$~\cite{Hashimoto:2006aw} & $26.8\pm0.8$ & $22.4\pm0.6$ & $17.3\pm0.7$ & $12.2\pm0.6$ & $7.1\pm0.6$ \\
\hline
\hline
\end{tabular}
\label{tab:LamExp}
\end{table*}

\section{Optimized parameters}
\label{app:OptParams}
\begin{figure*}[tbhp]
\centering
\includegraphics[width=5.5cm]{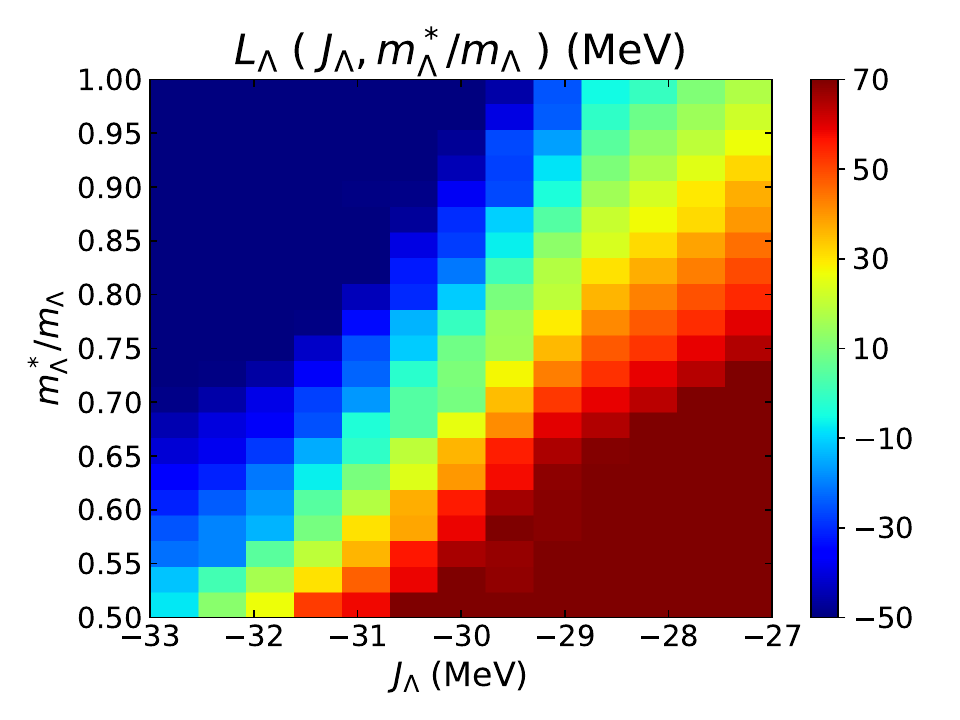}
\includegraphics[width=5.5cm]{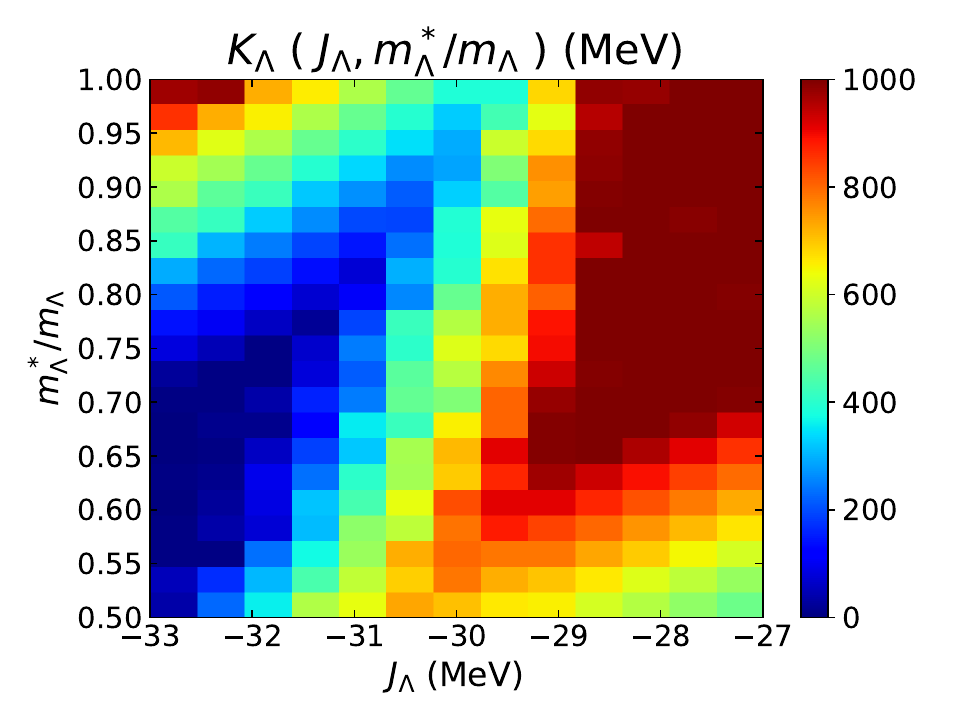}
\includegraphics[width=5.5cm]{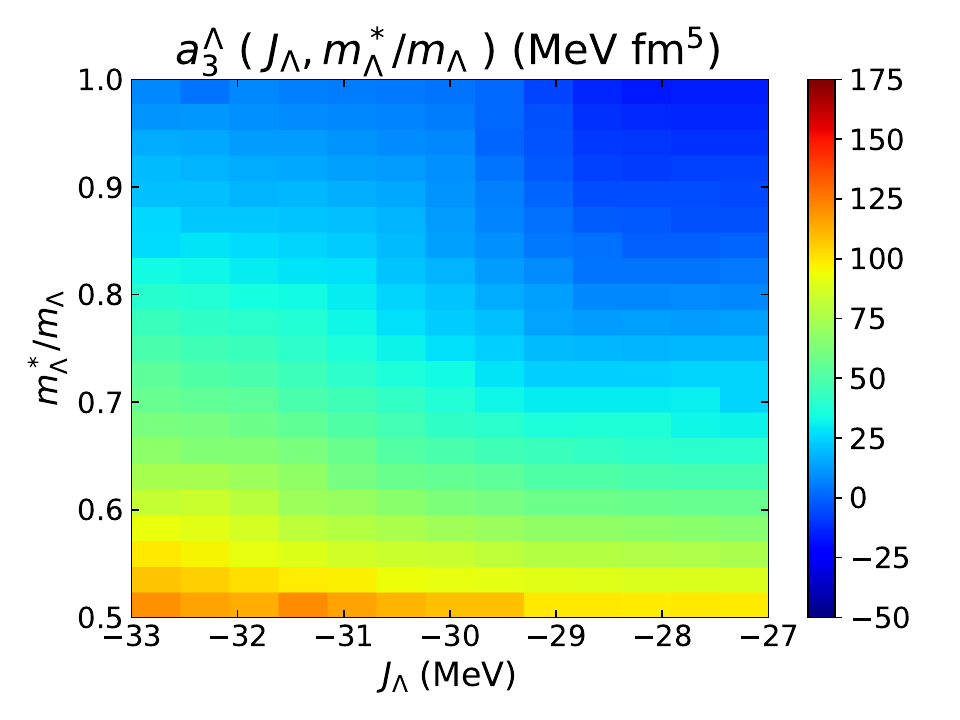}\\
\includegraphics[width=5.5cm]{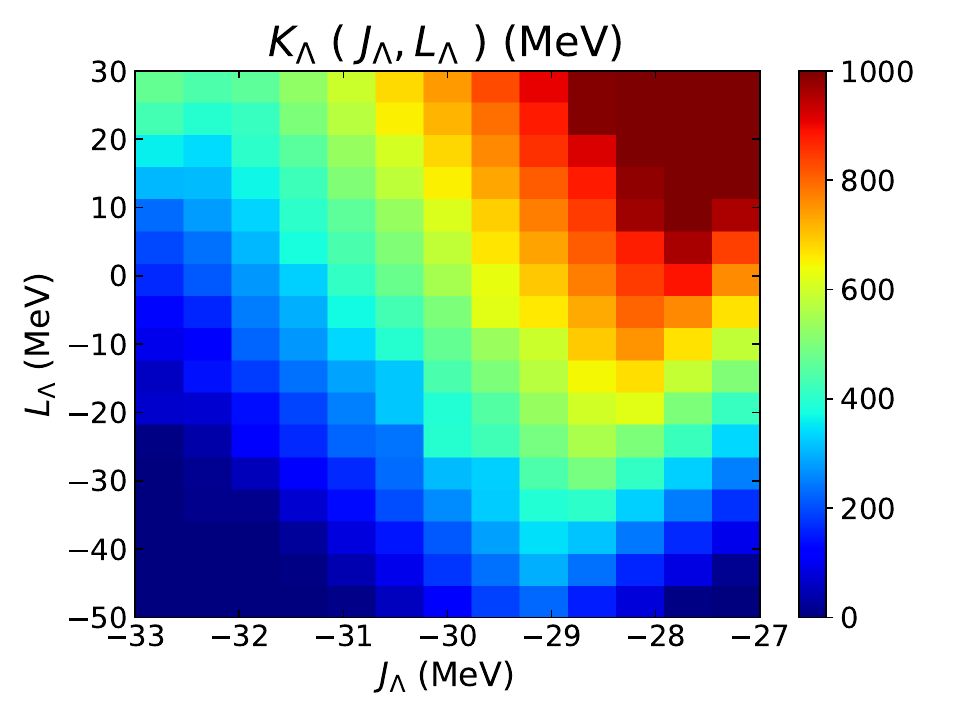}
\includegraphics[width=5.5cm]{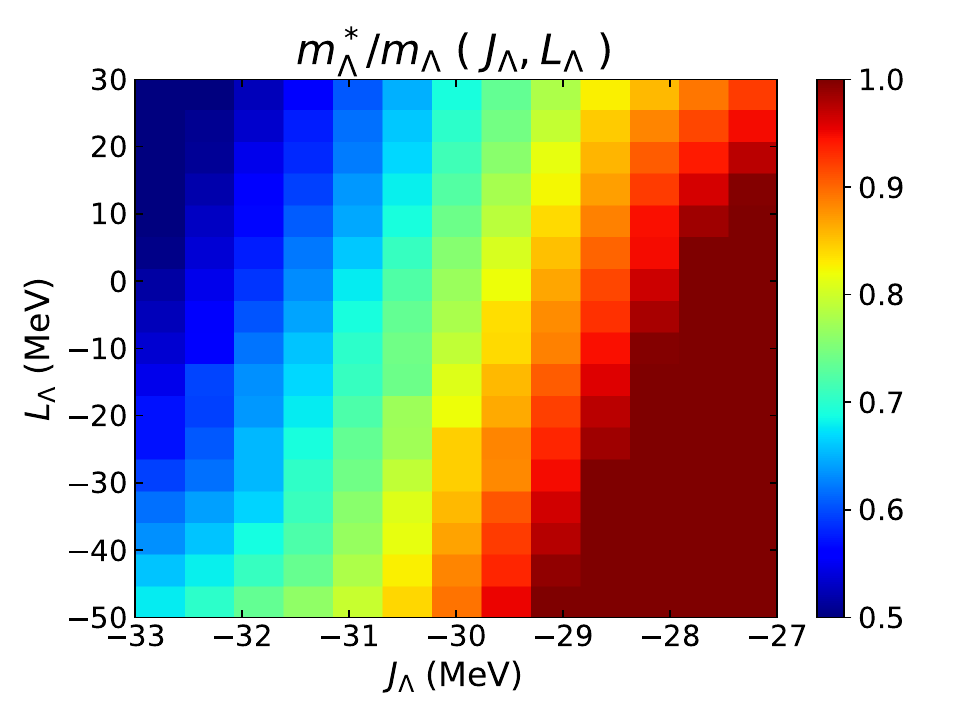}
\includegraphics[width=5.5cm]{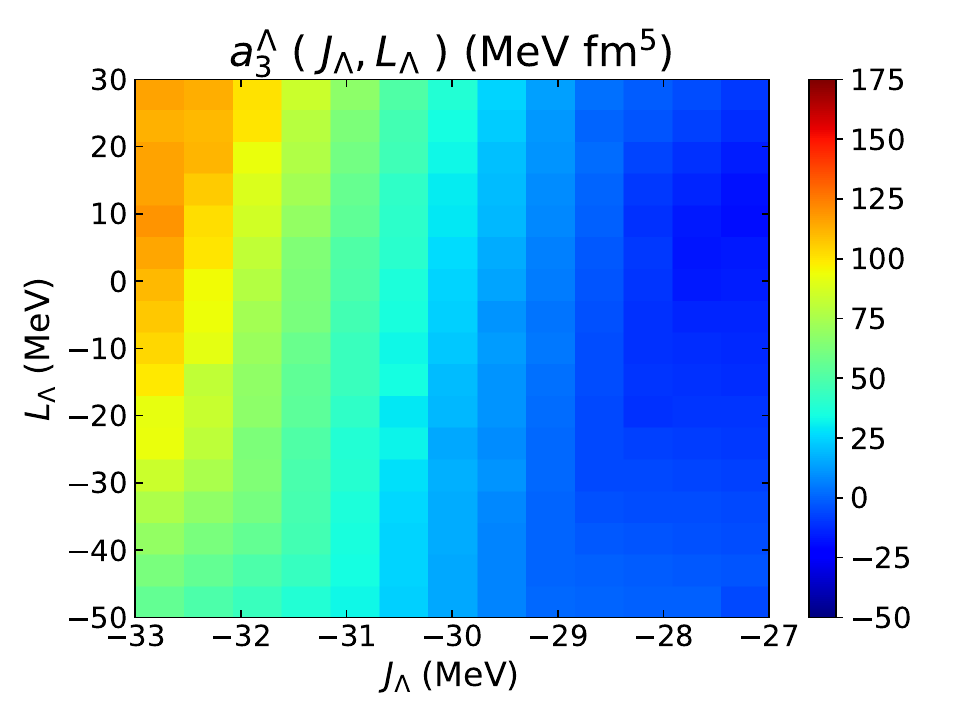}\\
\includegraphics[width=5.5cm]{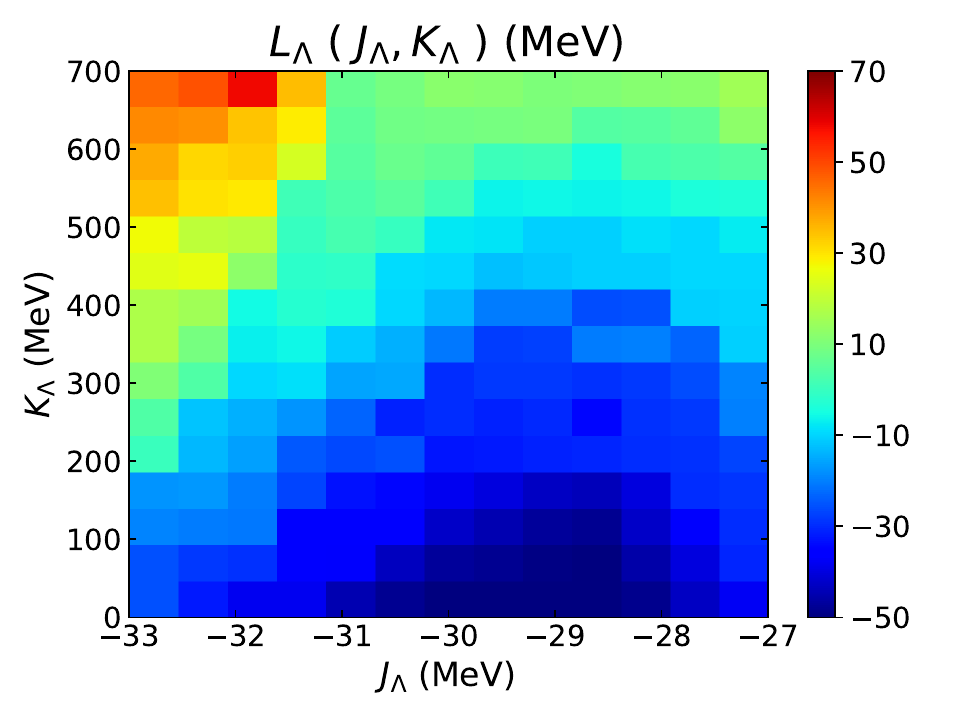}
\includegraphics[width=5.5cm]{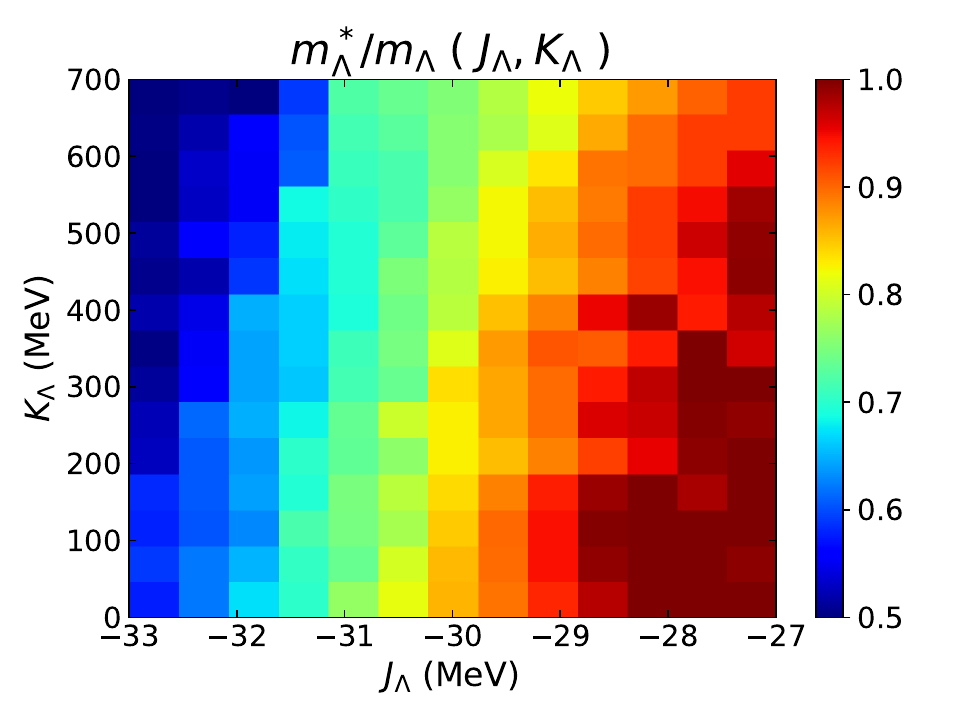}
\includegraphics[width=5.5cm]{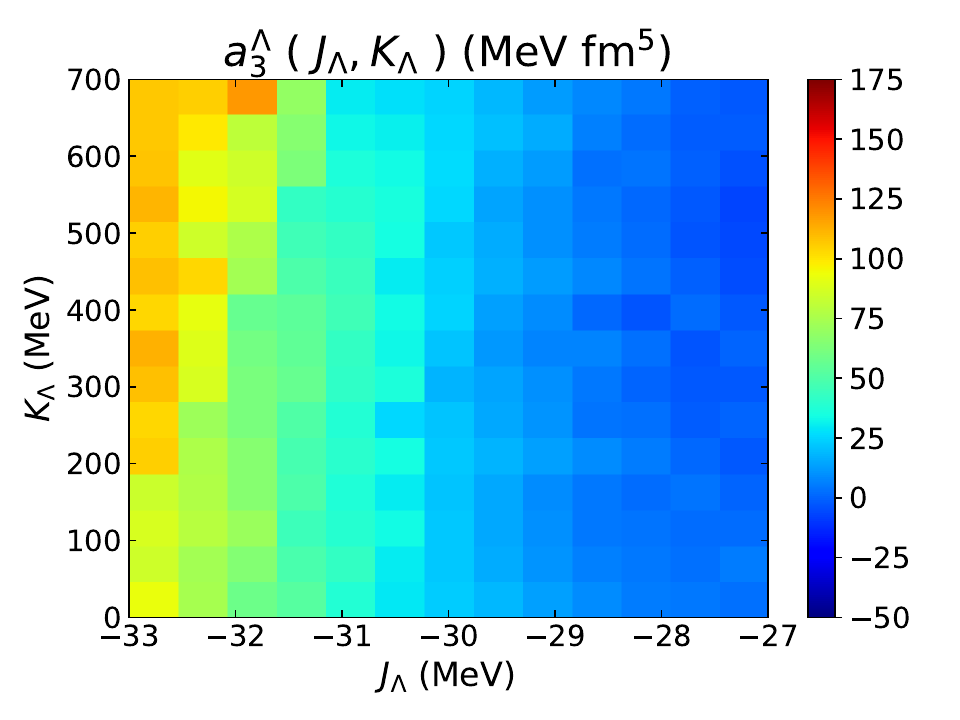}\\
\includegraphics[width=5.5cm]{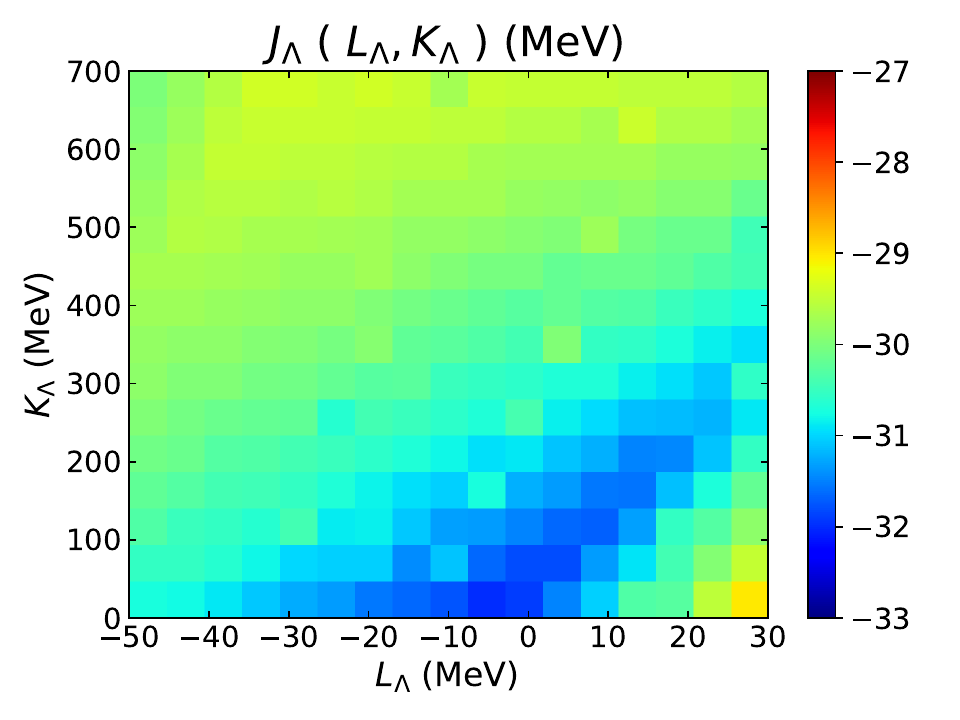}
\includegraphics[width=5.5cm]{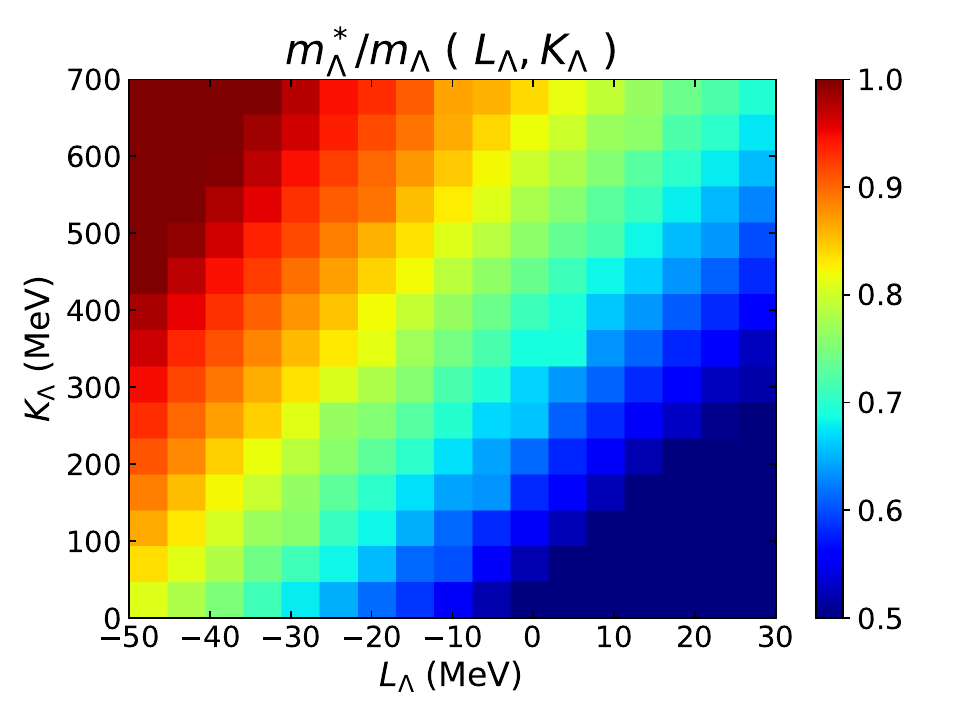}
\includegraphics[width=5.5cm]{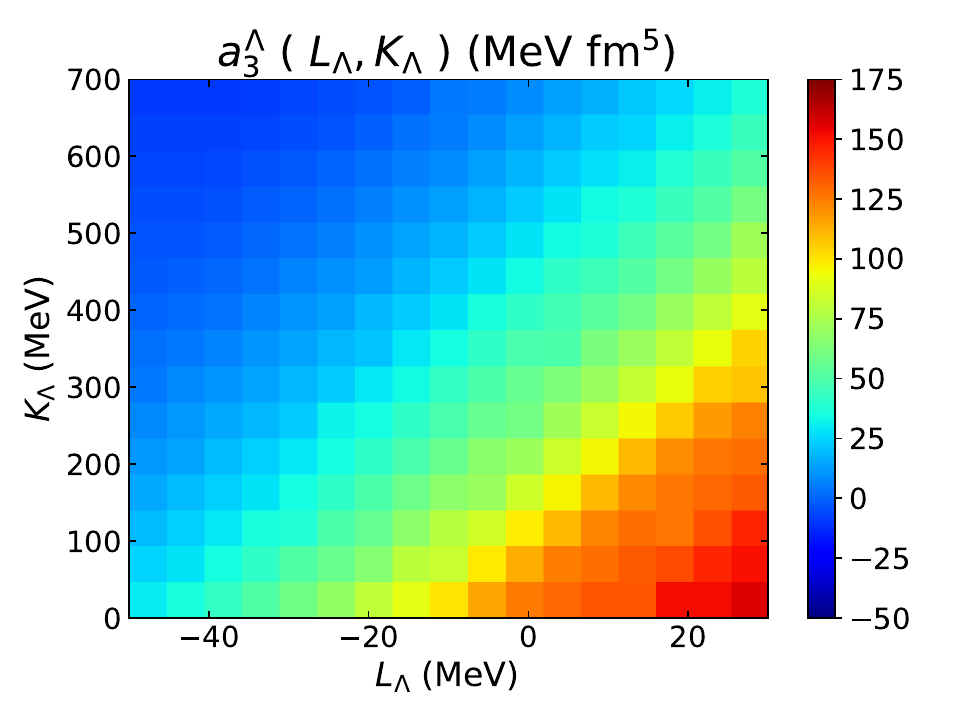}
\caption{
Heat maps of the optimized parameters in Fig. 6.
The parameters are optimized for minimizing $\Delta B_\Lambda$ at each grid point.
}
\label{fig:MissParam2dim}
\end{figure*}

\begin{figure*}[tbhp]
\centering
\includegraphics[width=7cm]{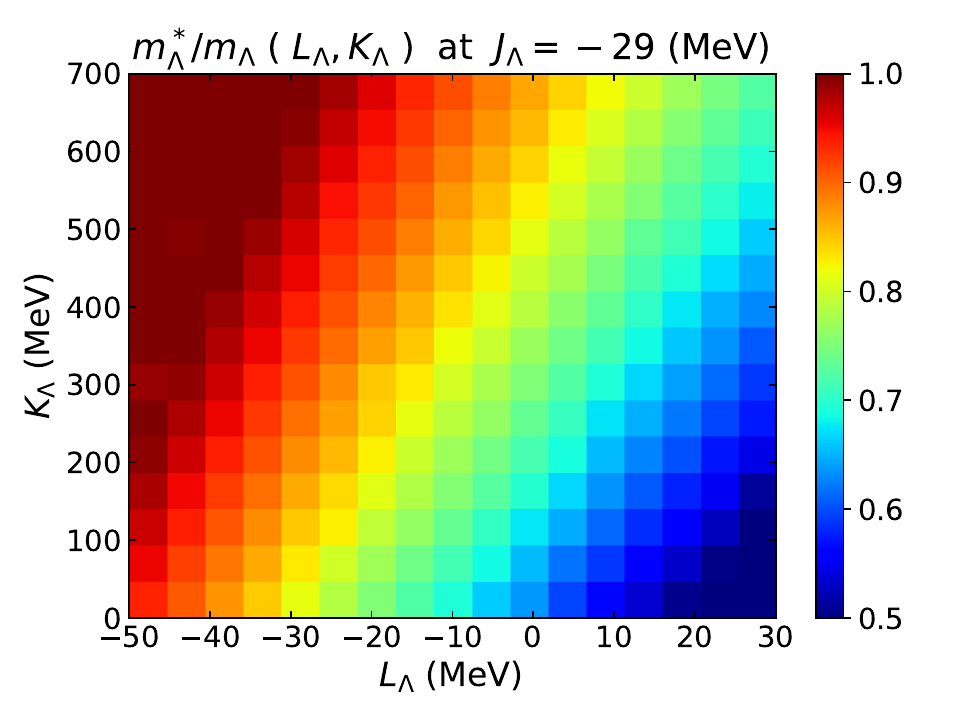}
\includegraphics[width=7cm]{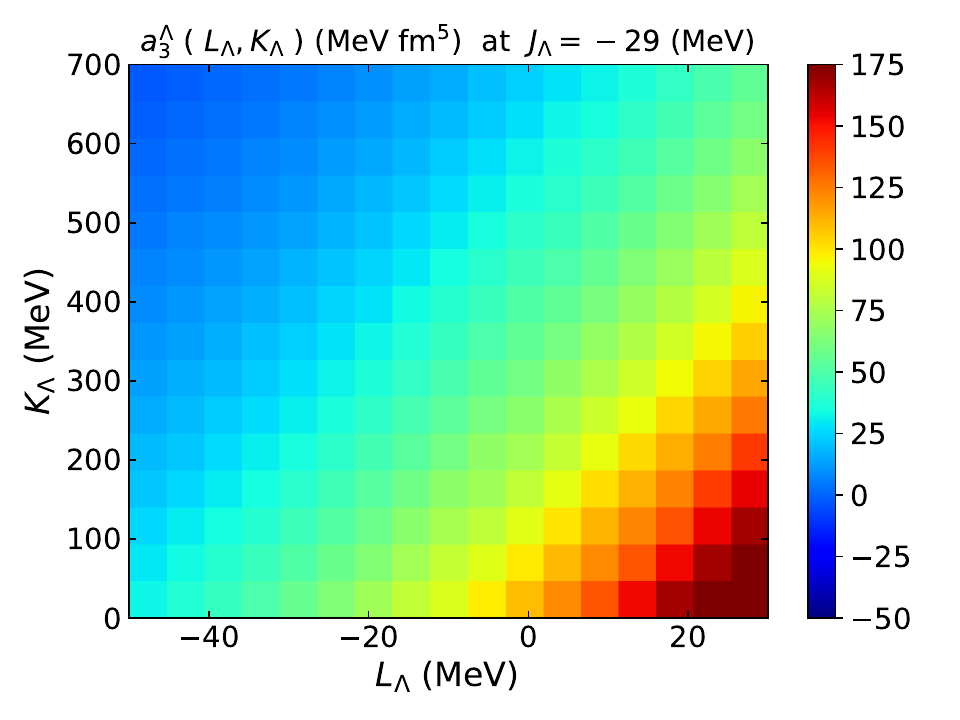}\\
\includegraphics[width=7cm]{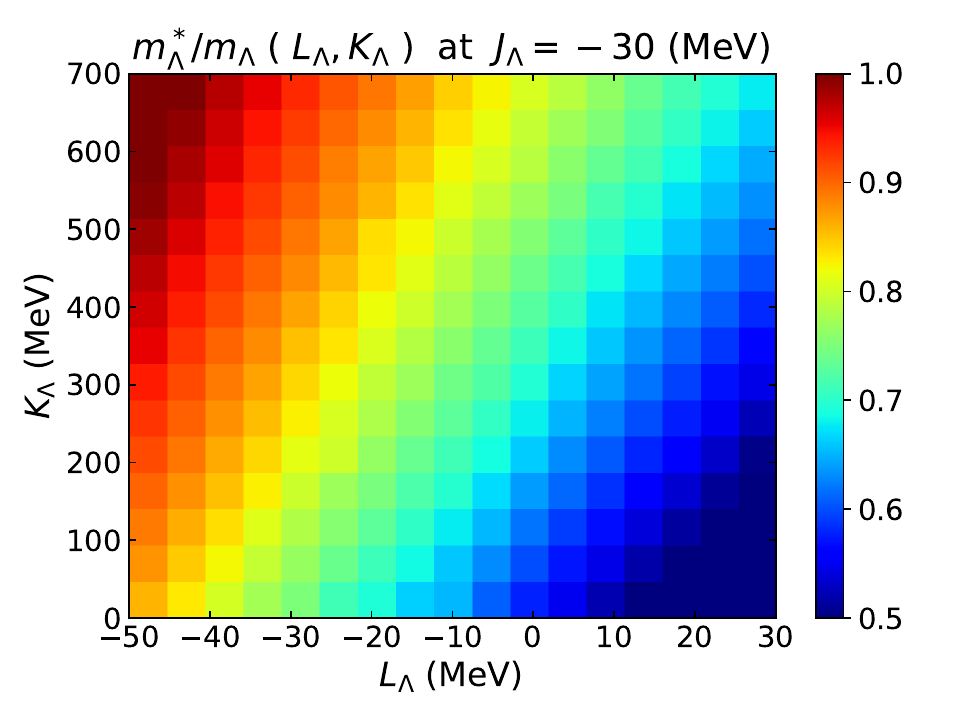}
\includegraphics[width=7cm]{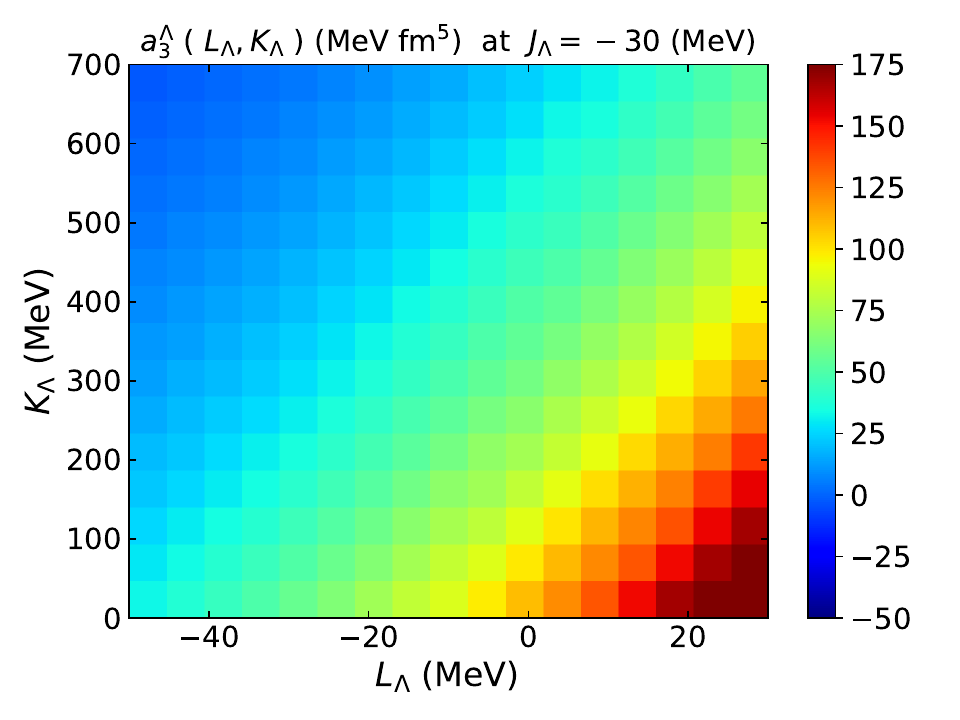}\\
\includegraphics[width=7cm]{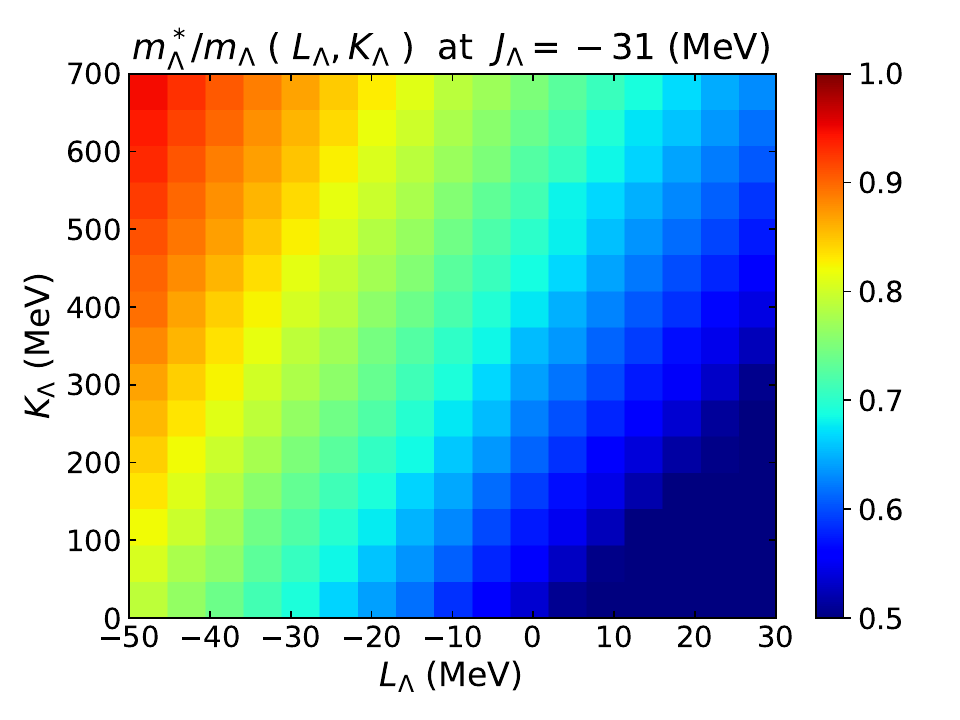}
\includegraphics[width=7cm]{a3_vsGivenLKJ30_GSS2.pdf}
\caption{Heat maps of the optimized parameters in Fig. 7.
}
\label{fig:MissParam1dim}
\end{figure*}

The optimized parameters in Figs.~\ref{fig:Jm_best} and \ref{fig:LK_J}
are summarized in Figs.~\ref{fig:MissParam2dim} and \ref{fig:MissParam1dim}, respectively. Those parameters are optimized for minimizing $\Delta B_\Lambda$ for each grid point.

\bibliography{ref}

\end{document}